\documentclass[12pt]{article}

\usepackage{amssymb}

\usepackage{amsmath, amssymb, amsfonts, amscd, xspace, pifont}
\usepackage{epsfig, psfrag, pstricks}
\usepackage{float, fancybox, fullpage}
\usepackage{harvard}
\usepackage{graphics}

\newcommand{\V}[1]{\ensuremath{\boldsymbol{#1}}\xspace}
\newcommand{\M}[1]{\ensuremath{\boldsymbol{#1}}\xspace}
\newcommand{\F}[1]{\ensuremath{\mathrm{#1}}\xspace}

\newcommand{\T}[1]{\ensuremath{{#1}^{\mbox{\sf\tiny T}}}}
\newtheorem{theorem}{Theorem}
\newtheorem{lemma}{Lemma}

\begin{document}
\setcounter{page}{1}

\title{Group Variable Selection via a Hierarchical Lasso and Its
 Oracle Property}
\author{Nengfeng Zhou\\
  Consumer Credit Risk Solutions\\
  Bank of America\\
  Charlotte, NC 28255\\
  \\
  Ji Zhu\footnote{Address for correspondence: Ji Zhu, 439 West
    Hall, 1085 South University Ave, Ann Arbor, MI
    48109-1107. E-mail: jizhu@umich.edu.}\\
  Department of Statistics\\
  University of Michigan\\
  Ann Arbor, MI 48109\\
}
\maketitle
\newpage

\begin{abstract}
In many engineering and scientific applications, prediction
variables are grouped, for example, in biological applications where
assayed genes or proteins can be grouped by biological roles or
biological pathways. Common statistical analysis methods such as
ANOVA, factor analysis, and functional modeling with basis sets also
exhibit natural variable groupings. Existing successful group
variable selection methods such as Antoniadis and Fan (2001), Yuan
and Lin (2006) and Zhao, Rocha and Yu (2009) have the limitation of
selecting variables in an ``all-in-all-out'' fashion, i.e., when one
variable in a group is selected, all other variables in the same
group are also selected. In many real problems, however, we may want
to keep the flexibility of selecting variables within a group, such
as in gene-set selection. In this paper, we develop a new group
variable selection method that not only removes unimportant groups
effectively, but also keeps the flexibility of selecting variables
within a group. We also show that the new method offers the
potential for achieving the theoretical ``oracle'' property as in
Fan and Li (2001) and Fan and Peng (2004).
\end{abstract}
\bigskip

\begin{quote} \small
{\bf Keywords}: Group selection; Lasso; Oracle property;
Regularization; Variable selection
\end{quote} \normalsize

\newpage
\setcounter{page}{1}

\section{Introduction} \label{sec:intro}

Consider the usual regression situation: we have training data,
$(\V{x}_1,y_1)$, $\ldots$, $(\V{x}_i,y_i)$, $\ldots$,
$(\V{x}_n,y_n)$, where $\V{x}_i = (x_{i1},\ldots,x_{ip})$ are the
predictors and $y_i$ is the response. To model the response $y$ in
terms of the predictors $x_1,\ldots,x_p$, one may consider the
linear model:
\begin{equation} \label{intro:eq01}
 y = \beta_0 + \beta_1 x_{1} + \ldots + \beta_p x_{p} + \varepsilon,
\end{equation}
where $ \varepsilon$ is the error term. In many important practical
problems, however, prediction variables are ``grouped.'' For example,
in ANOVA factor analysis, a factor may have several levels and can
be expressed via several dummy variables, then the dummy variables
corresponding to the same factor form a natural ``group.''
Similarly, in additive models, each original prediction variable may
be expanded into different order polynomials or a set of basis
functions, then these polynomials (or basis functions) corresponding
to the same original prediction variable form a natural ``group.''
Another example is in biological applications, where assayed genes
or proteins can be grouped by biological roles (or biological
pathways).

For the rest of the paper, we assume that the prediction variables
can be divided into $K$ groups and the $k$th group contains $p_k$
variables. Specifically, the linear model (\ref{intro:eq01}) is now
written as
\begin{eqnarray} \label{intro:eq02}
 {y}_i &= &
 \beta_0 + \sum_{k=1}^{K} \sum_{j=1}^{p_k} \beta_{kj} x_{i,kj} +
 \varepsilon_i.
\end{eqnarray}
And we are interested in finding out which variables, especially
which ``groups,'' have an important effect on the response. For
example, $(x_{11},\ldots,x_{1p_1})$, $(x_{21},\ldots,x_{2p_2})$,
$\ldots$, $(x_{K1},\ldots,x_{Kp_K})$ may represent different
biological pathways, $y$ may represent a certain phenotype and we
are interested in deciphering which and how these biological
pathways ``work together'' to affect the phenotype.

There are two important challenges in this problem: prediction
accuracy and interpretation. We would like our model to accurately
predict on future data. Prediction accuracy can often be improved by
shrinking the regression coefficients. Shrinkage sacrifices some
bias to reduce the variance of the predicted value and hence may
improve the overall prediction accuracy. Interpretability is often
realized via variable selection. With a large number of prediction
variables,
we often would like to determine a smaller subset that exhibits the
strongest effects.

Variable selection has been studied extensively in the literature,
for example, see \citeasnoun{GeorgeMcCulloch93},
\citeasnoun{Breiman95}, \citeasnoun{Tibshirani96},
\citeasnoun{GeorgeFoster00}, \citeasnoun{FanLi01},
\citeasnoun{ZouHastie05},
\citeasnoun{LinZhang06} and \citeasnoun{WuEtAl07}.
In particular, lasso \cite{Tibshirani96}
has gained much attention in recent years. The lasso criterion
penalizes the $L_1$-norm of the regression coefficients to achieve a
sparse model:
\begin{equation} \label{eq:crit03}
 \max_{\beta_0, \beta_{kj}} - \frac{1}{2} \sum_{i=1}^{n} \left( y_i
 -\beta_0 - \sum_{k=1}^{K} \sum_{j=1}^{p_k}
 \beta_{kj} x_{i,kj} \right)^2 - \lambda \sum_{k=1}^{K}
 \sum_{j=1}^{p_k} |\beta_{kj}|,
\end{equation}
where $\lambda \geq 0$ is a tuning parameter. Note that by location
transformation, we can always assume that the predictors and the
response have mean 0, so we can ignore the intercept in equation
(\ref{eq:crit03}).

Due to the singularity at $\beta_{kj} = 0$, the $L_1$-norm penalty
can shrink some of the fitted coefficients to be {\it exact} zero
when making the tuning parameter sufficiently large. However, lasso
and other methods above are for the case when the candidate
variables can be treated individually or ``flatly.'' When variables
are grouped, ignoring the group structure and directly applying
lasso as in (\ref{eq:crit03}) may be sub-optimal. For example,
suppose the $k$th group is unimportant, then lasso tends to make
individual estimated coefficients in the $k$th group to be zero,
rather than the whole group to be zero, i.e., lasso tends to make
selection based on the strength of individual variables rather than
the strength of the group, often resulting in selecting more groups
than necessary.

\citeasnoun{AntoniadisFan01}, \citeasnoun{YuanLin06} and
\citeasnoun{ZhaoRochaYu06} have addressed the group variable
selection problem in the literature. \citeasnoun{AntoniadisFan01}
proposed to use a
blockwise additive penalty in the setting of wavelet approximations.
To increase the estimation precision, empirical wavelet
coefficients were thresholded or shrunken in blocks (or groups)
rather than individually.

\citeasnoun{YuanLin06} and \citeasnoun{ZhaoRochaYu06}
extended the lasso model (\ref{eq:crit03}) for group variable
selection.
\citeasnoun{YuanLin06} chose to penalize the $L_2$-norm of the
coefficients within each group, i.e., $\sum_{k=1}^K \|
\V{\beta}_{k} \|_2$, where
\begin{equation} \label{intro:eq05}
  \| \V{\beta}_{k} \|_2 =
  \sqrt{\beta_{k1}^2 + \ldots + \beta_{kp_k}^2}.
\end{equation}
Due to the singularity of $\| \V{\beta}_{k} \|_2$ at $\V{\beta}_k =
\V{0} $, appropriately tuning $\lambda$ can set the whole
coefficient vector $\V{\beta}_k = \V{0}$, hence the $k$th group is
removed from the fitted model.  We note that in the setting of
wavelet analysis, this method reduces to
\citeasnoun{AntoniadisFan01}.

Instead of using the $L_2$-norm penalty, \citeasnoun{ZhaoRochaYu06}
suggested using the $L_{\infty}$-norm penalty, i.e.,
$\sum_{k=1}^K \| \V{\beta}_{k} \|_{\infty}$, where
\begin{equation} \label{intro:eq07}
  \| \V{\beta}_{k} \|_{\infty} = \max ( {|\beta_{k1}|,
    |\beta_{k2}|, \ldots, |\beta_{kp_k}| } ).
\end{equation}
Similar to the $L_2$-norm, the $L_{\infty}$-norm of $\V{\beta}_{k}$
is also singular when $\V{\beta}_k = \V{0} $; hence when $\lambda$
is appropriately tuned, the $L_{\infty}$-norm can also effectively
remove unimportant groups.

However, there are some possible limitations with these methods:
Both the $L_2$-norm penalty and the $L_{\infty}$-norm penalty select
variables in an ``all-in-all-out'' fashion, i.e., when one variable
in a group is selected, {\it all} other variables in the same group
are also selected. The reason is that both $ \| \V{\beta}_{k} \|_2$
and $\| \V{\beta}_{k} \|_{\infty}$ are singular only when the whole
vector $\V{\beta}_k = \V{0}$. Once a component of $\V{\beta}_k$ is
non-zero, the two norm functions are no longer singular. This can
also be heuristically understood as the following: for the
$L_2$-norm (\ref{intro:eq05}), it is the ridge penalty that is under
the square root; since the ridge penalty can not do variable
selection (as in ridge regression), once the $L_2$-norm is non-zero
(or the corresponding group is selected), all components will be
non-zero. For the $L_{\infty}$-norm (\ref{intro:eq07}), if the
``max($\cdot$)'' is non-zero, there is no increase in the penalty
for letting all the individual components move away from zero. Hence
if one variable in a group is selected, all other variables are also
automatically selected.

In many important real problems, however, we may want to keep the
flexibility of selecting variables {\it within} a group. For
example, in the gene-set selection problem, a biological pathway may
be related to a certain biological process, but it does not
necessarily mean all the genes in the pathway are all related to the
biological process. We may want to not only remove unimportant
pathways effectively, but also identify important genes within
important pathways.

For the $L_{\infty}$-norm penalty, another possible limitation is
that the estimated coefficients within a group tend to have the same
magnitude, i.e. $|\beta_{k1}| = |\beta_{k2}| = \ldots =
|\beta_{kp_k}|$; and this may cause some serious bias, which
jeopardizes the prediction accuracy.

In this paper, we propose an extension of lasso for group variable
selection, which we call hierarchical lasso (HLasso). Our method not
only removes unimportant groups effectively, but also keeps the
flexibility of selecting variables within a group. Furthermore,
asymptotic studies motivate us to improve our model and show that
when the tuning parameter is appropriately chosen, the improved
model has the {\it oracle} property \cite{FanLi01,FanPeng04}, i.e., it
performs as well as if the correct underlying model were given in
advance. Such a theoretical property has not been previously studied
for group variable selection at both the group level and the within
group level.

The rest of the paper is organized as follows. In Section
\ref{sec:model}, we introduce our new method: the hierarchical
lasso. We propose an algorithm to compute the solution for the
hierarchical lasso in Section \ref{sec:alg}. In Sections
\ref{sec:theory} and \ref{sec:improve}, we study the asymptotic
behavior of the hierarchical lasso and propose an improvement for
the hierarchical lasso.
Numerical results are in
Sections \ref{sec:resultsimu} and \ref{sec:resultreal}, and we
conclude the paper with Section \ref{sec:summary}.

\section{Hierarchical Lasso}
\label{sec:model}

In this section, we extend the lasso method for group variable
selection so that we can effectively remove unimportant variables at
both the group level and the within group level.

We reparameterize $\beta_{kj}$ as
\begin{equation} \label{intro:eq08}
 \beta_{kj} = d_k \alpha_{kj} , ~~~ k = 1,\ldots,K; ~j =
 1,\ldots,p_k,
\end{equation}
where $d_k \ge 0 $ (for identifiability reasons). This decomposition
reflects the information that $\beta_{kj}, j = 1, \ldots, p_k$, all
belong to the $k$h group, by treating each $\beta_{kj}$
hierarchically. $d_k$ is at the first level of the hierarchy,
controlling $\beta_{kj}, j = 1, \ldots, p_k$, as a group;
$\alpha_{kj}$'s are at the second level of the hierarchy, reflecting
differences within the $k$th group.

For the purpose of variable selection, we consider the following
penalized least squares criterion:

\begin{eqnarray}\label{model:eq01}
 \max_{d_k,\alpha_{kj} } && - \frac{1}{2} \sum_{i=1}^{n} \left(
 y_{i} - \sum_{k=1}^K d_k\sum_{j=1}^{p_k} \alpha_{kj} x_{i,kj}
 \right)^2 \nonumber \\
 && - \lambda_1 \cdot \sum_{k=1}^K d_k
 - \lambda_2 \cdot \sum_{k=1}^K \sum_{j=1}^{p_k} |\alpha_{kj}| \\
 \nonumber
 \textrm{subject to} && d_k\ge 0, ~k=1, \ldots, K,
\end{eqnarray}
where $\lambda_1 \ge 0 $ and $\lambda_2 \ge 0$ are tuning
parameters. $\lambda_1$ controls the estimates at the group level,
and it can effectively remove unimportant groups: if $d_k$ is
shrunken to zero, all $\beta_{kj}$ in the $k$th group will be equal
to zero. $\lambda_2$ controls the estimates at the variable-specific
level: if $d_k$ is not equal to zero, some of the $\alpha_{kj}$
hence some of the $\beta_{kj},j=1,\ldots,p_k$, still have the
possibility of being zero; in this sense, the hierarchical penalty
keeps the flexibility of the $L_1$-norm penalty.

One may complain that such a hierarchical penalty may be more
complicated to tune in practice, however, it turns out that the two
tuning parameters $\lambda_1$ and $\lambda_2$ in (\ref{model:eq01})
can be simplified into one. Specifically, let $\lambda = \lambda_1
\cdot \lambda_2$, we can show that (\ref{model:eq01}) is equivalent
to
\begin{eqnarray} \label{model:eq02}
 \max_{d_k,\alpha_{kj} } && - \frac{1}{2} \sum_{i=1}^{n} \left(
 y_{i} - \sum_{k=1}^K d_k\sum_{j=1}^{p_k} \alpha_{kj}
 x_{i,kj} \right)^2
 - \sum_{k=1}^K d_k - \lambda \sum_{k=1}^K \sum_{j=1}^{p_k} |\alpha_{kj}|
 \\ \nonumber
 \textrm{subject to} && d_k\ge 0, k=1, \ldots, K.
\end{eqnarray}

\begin{lemma} \label{lemma0}
 Let ($\hat{\V{d}}^{\ast}, \hat{\V{\alpha}}^{\ast}$) be a local maximizer of (\ref{model:eq01}),
 then there exists a local maximizer ($\hat{\V{d}}^{\star},
 \hat{\V{\alpha}}^{\star}$) of (\ref{model:eq02}) such that
 $ \hat{d}_k^{\ast} \hat{\alpha}_{kj}^{\ast} = \hat{d}_k^\star
 \hat{\alpha}_{kj}^\star. $
 Similarly, if ($\hat{\V{d}}^{\star}, \hat{\V{\alpha}}^{\star}$)
 is a local maximizer of (\ref{model:eq02}),
 there exists a local maximizer ($\hat{\V{d}}^{\ast},
 \hat{\V{\alpha}}^{\ast}$) of (\ref{model:eq01})
 such that
 $ \hat{d}_k^{\ast} \hat{\alpha}_{kj}^{\ast} = \hat{d}_k^\star
 \hat{\alpha}_{kj}^\star. $
\end{lemma}

The proof is in the Appendix. This lemma indicates that the final
fitted models from (\ref{model:eq01}) and (\ref{model:eq02}) are the
same, although they may provide different $d_k$ and $\alpha_{kj}$.
This also implies that in practice, we do not need to tune
$\lambda_1$ and $\lambda_2$ separately; we only need to tune one
parameter $\lambda = \lambda_1\cdot\lambda_2$ as in
(\ref{model:eq02}).

\section{Algorithm}
\label{sec:alg}

To estimate the
$d_k$ and $\alpha_{kj}$ in (\ref{model:eq02}), we can use an
iterative approach, i.e., we first fix $d_k$ and estimate
$\alpha_{kj}$, then we fix $\alpha_{kj}$ and estimate $d_k$, and we
iterate between these two steps until the solution converges. Since
at each step, the value of the objective function (\ref{model:eq02})
decreases, the solution is guaranteed to converge.

When $d_k$ is fixed, (\ref{model:eq02}) becomes a lasso problem,
hence we can use either the LAR/LASSO algorithm \cite{Efron04} or a
quadratic programming package to efficiently solve for
$\alpha_{kj}$. When $\alpha_{kj}$ is fixed, (\ref{model:eq02})
becomes a non-negative garrote problem. Again, we can use either an
efficient solution path algorithm or a quadratic programming package
to solve for $d_k$. In summary, the algorithm proceeds as follows:

\begin{itemize}
\item[1.] (Standardization) Center $\V{y}$. Center and normalize $\V{x}_{kj}$.
\item[2.] (Initialization) Initialize $d_k^{(0)}$ and
 $\alpha_{kj}^{(0)}$ with some plausible values.
 For example, we can set $d_k^{(0)}=1$ and use the least squares
 estimates or the simple
 regression estimates by regressing the response $\V{y}$ on each
 of the $\V{x}_{kj}$ for
 $\alpha_{kj}^{(0)}$. Let $ \beta_{kj}^{(0)} = d_{k}^{(0)}
 \alpha_{kj}^{(0)}$ and $m$ = 1.
\item[3.] (Update $\alpha_{kj}$) Let
 \[ \tilde{x}_{i,kj} = d_k^{(m-1)} x_{i,kj}, ~~ k = 1,\ldots,K;
 ~j = 1,\ldots, p_k, \]
 then
 \[ \alpha_{kj}^{(m)} = \arg \max_{\alpha_{kj}}
 - \frac{1}{2} \sum_{i=1}^{n} \left( y_{i} - \sum_{k=1}^K
 \sum_{j=1}^{p_k} \alpha_{kj} \tilde{x}_{i,kj} \right)^2
 - \lambda \sum_{k=1}^K \sum_{j=1}^{p_k} |\alpha_{kj}|. \]
\item[4.] (Update $d_{k}$) Let
 \[ \tilde{x}_{i,k} = \sum_{j=1}^{p_k} \alpha_{kj}^{(m)}
 x_{i,kj}, ~~ k = 1,\ldots,K, \]
 then
 \[ d_{k}^{(m)} = \arg \max_{d_{k} \ge 0}
 - \frac{1}{2} \sum_{i=1}^{n} \left( y_{i} - \sum_{k=1}^K
 d_{k} \tilde{x}_{i,k} \right)^2
 - \sum_{k=1}^K d_{k}. \]
\item[5.] (Update $\beta_{kj}$) Let
 \[ \beta_{kj}^{(m)} = d_{k}^{(m)} \alpha_{kj}^{(m)}. \]
\item[6.] If $\| \beta_{kj}^{(m)} - \beta_{kj}^{(m-1)} \|$ is
 small enough, stop the algorithm. Otherwise, let $m \leftarrow m+1$ and
 go back to Step 3.
\end{itemize}

\subsection{Orthogonal Case}

To gain more insight into the hierarchical penalty, we have also
studied the algorithm in the orthogonal design case. This can be
useful, for example, in the wavelet setting, where each $\V{x}_{kj}$
corresponds to a wavelet basis function, different $k$ may
correspond to different ``frequency'' scales, and different $j$ with
the same $k$ correspond to different ``time'' locations.
Specifically, suppose $\T{\V{x}}_{kj} \V{x}_{kj} = 1$ and
$\T{\V{x}}_{kj} \V{x}_{k^{\prime}j^{\prime}} = 0$ if $k\neq
k^{\prime}$ or $j\neq j^{\prime}$, then Step 3 and Step 4 in the
above algorithm have closed form solutions.

Let $\hat{\beta}_{kj}^{\mbox{ols}} = \T{\V{x}}_{kj}\V{y}$ be the ordinary
least squares solution when $\V{x}_{kj}$ are orthonormal to each
other.
\begin{itemize}
\item[Step 3.] When $d_k$ is fixed,
 \begin{equation} \label{orth:eq01}
 \alpha_{kj}^{(m)} = \mathbb{I}(d_k^{(m-1)}>0) \cdot
 \textrm{sgn}( \hat{\beta}_{kj}^{\mbox{ols}} ) \cdot \left( \frac{|
 \hat{\beta}_{kj}^{\mbox{ols}}|}{d_k^{(m-1)} } -
 \frac{\lambda}{(d_k^{(m-1)})^2} \right)_+ .
 \end{equation}
\item[Step 4.] When $\alpha_{kj}$ is fixed,
 \begin{equation} \label{orth:eq02}
 d_k^{(m)} = \mathbb{I}(\exists j, \alpha_{kj}^{(m)} \neq 0) \cdot
 \left( \sum_{j=1}^{p_k}
 \frac{(\alpha_{kj}^{(m)})^2}{\sum_{j=1}^{p_k} (\alpha_{kj}^{(m)})^2}
 \frac{\hat{\beta}_{kj}^{\mbox{ols}}}{\alpha_{kj}^{(m)}} -
 \frac{1}{\sum_{j=1}^{p_k} (\alpha_{kj}^{(m)})^2} \right)_+ .
 \end{equation}
\end{itemize}
Equations (\ref{orth:eq01}) and (\ref{orth:eq02}) show that both
$d_k^{(m)}$ and $\alpha_{kj}^{(m)}$ are soft-thresholding estimates.
Here we provide some intuitive explanation.

We first look at $\alpha_{kj}^{(m)}$ in equation (\ref{orth:eq01}).
If $d_{k}^{(m-1)} = 0$, it is natural to estimate all $\alpha_{kj}$ to
be zero because of the penalty on $\alpha_{kj}$. If $d_{k}^{(m-1)} >
0$, then from our reparametrization, we have $\alpha_{kj} =
\beta_{kj} / d_k^{(m-1)} $, $j = 1, \ldots,p_k$. Plugging in
$\hat{\beta}_{kj}^{\mbox{ols}}$ for $\beta_{kj}$, we obtain
$\tilde{\alpha}_{kj} = \hat{\beta}_{kj}^{\mbox{ols}} / d_k^{(m-1)}$.
Equation (\ref{orth:eq01}) shrinks $\tilde{\alpha}_{kj}$, and the
amount of shrinkage is inversely proportional to $(d_k^{(m-1)})^2$.
So when $ d_k^{(m-1)} $ is large, which indicates the $k$th group is
important, the amount of shrinkage is small, while when $
d_k^{(m-1)} $ is small, which indicates the $k$th group is less
important, the amount of shrinkage is large.

Now considering $d_k^{(m)}$ in equation (\ref{orth:eq02}). If all
$\alpha_{kj}^{(m)}$ are zero, it is natural to estimate $d_k^{(m)}$
also to be zero because of the penalty on $d_k$. If not all
$\alpha_{kj}^{(m)}$ are 0, say $\alpha_{kj_1}^{(m)}, \ldots,
\alpha_{kj_r}^{(m)} $ are not zero, then we have $d_k = \beta_{kj_s}
/\alpha_{kj_s}^{(m)}, 1\le s \le r$. Again, plugging in
$\hat{\beta}_{kj_s}^{\mbox{ols}}$ for $\beta_{kj_s}$, we obtain $r$
estimates for $d_k$: $\tilde{d}_k = \hat{\beta}_{kj_s}^{\mbox{ols}}
/\alpha_{kj_s}^{(m)}, 1\le s \le r$. A natural estimate for $d_k$ is
then a weighted average of the $\tilde{d}_k$, and equation
(\ref{orth:eq02}) provides such a (shrunken) average, with weights
proportional to $(\alpha_{kj}^{(m)})^2$.

\section{Asymptotic Theory}
\label{sec:theory}

In this section, we explore the asymptotic behavior of the
hierarchical lasso method.

The hierarchical lasso criterion (\ref{model:eq02}) uses $d_k$ and
$\alpha_{kj}$. We first show that it can also be written in an
equivalent form using the original regression coefficients
$\beta_{kj}$.


\begin{theorem} \label{thm1}
If ($\hat{\V{d}}, \hat{\V{\alpha}}$) is a
local maximizer of (\ref{model:eq02}), then $\hat{\V{\beta}}$, where
$\hat{\beta}_{kj} = \hat{d}_k \hat{\alpha}_{kj} $, is a local
maximizer of
\begin{eqnarray} \label{asym:eq01}
  \max_{\beta_{kj} } && - \frac{1}{2}
  \sum_{i=1}^{n} \left( y_{i} - \sum_{k=1}^K \sum_{j=1}^{p_k}
  x_{i,kj}\beta_{kj} \right)^2 \nonumber \\
  & & - 2\sqrt{\lambda} \cdot \sum_{k=1}^K
  \sqrt{ |\beta_{k1}|
    + |\beta_{k2}| + \ldots + |\beta_{kp_k}|}.
\end{eqnarray}
On the other hand, if $\hat{\V{\beta}}$ is a local maximizer of
(\ref{asym:eq01}), then
we define ($\hat{\V{d}}, \hat{\V{\alpha}}$), where
$\hat{d}_k = 0, \hat{\V{\alpha}}_k =0 $ if $\|
\hat{\V{\beta}}_{k} \|_1 = 0$,
and $\hat{d}_k = \sqrt{ \lambda \| \hat{\V{\beta}}_{k}
  \|_1 }, \hat{\V{\alpha}}_k = \frac{ \hat{\V{\beta}}_k }{
  \sqrt{\lambda \| \hat{\V{\beta}}_{k} \|_1 } }$
if $\| \hat{\V{\beta}}_{k} \|_1 \ne 0$.
Then the so-defined ($\hat{\V{d}}, \hat{\V{\alpha}}$) is a local
maximizer of (\ref{model:eq02}).
\end{theorem}

Note that the penalty term in (\ref{asym:eq01}) is similar to the
$L_2$-norm penalty (\ref{intro:eq05}), except that under each
square root, we now penalize the $L_1$-norm of $\V{\beta}_k$,
rather than the sum of squares. However, unlike the $L_2$-norm,
which is singular
only at the point $\V{\beta}_k = \V{0}$, (i.e., the whole vector is
equal to $\V{0}$), the square root of the $L_1$-norm is singular at
$\beta_{kj} = 0$ no matter what are the values of other
$\beta_{kj}$'s. This explains, from a different perspective, why the
hierarchical lasso can remove not only groups, but also variables
within a group even when the group is selected. Equation
(\ref{asym:eq01}) also implies that the hierarchical lasso belongs
to the ``CAP'' family in \citeasnoun{ZhaoRochaYu06}.
%

{ We study the asymptotic properties allowing the total number of
variables $P_n$, as well as the number of groups $K_n$ and the
number of
  variables within each group $p_{nk}$, to go to $\infty$, where
  $P_n = \sum_{k=1}^{K_n} p_{nk}$.
  Note that we add a subscript ``$n$'' to $K$ and $p_k$ to
  denote that these quantities can change with $n$.
Accordingly, $\V{\beta}$, $y_i$ and $x_{i,kj}$ are also changed to
$\V{\beta}_n$, $y_{ni}$ and $x_{ni,kj}$.} We write $2\sqrt{\lambda}$
in (\ref{asym:eq01}) as $n\lambda_n$, and the criterion
(\ref{asym:eq01}) is re-written as
\begin{eqnarray} \label{asym:eq01n}
  \max_{\beta_{n,kj} } && - \frac{1}{2}
  \sum_{i=1}^{n} \left( y_{ni} - \sum_{k=1}^{K_n} \sum_{j=1}^{p_{nk}}
  x_{ni,kj}\beta_{n,kj} \right)^2 \nonumber \\
  & & - n{\lambda_n} \cdot \sum_{k=1}^{K_n}
  \sqrt{ |\beta_{n,k1}|
    + |\beta_{n,k2}| + \ldots + |\beta_{n,kp_{nk}}|}.
\end{eqnarray}
Our asymptotic analysis in this section is based on the criterion
(\ref{asym:eq01n}).

 Let $\V{\beta}_n^0 = \T{(\V{\beta}_{{\cal
A}_n}^0, \V{\beta}_{{\cal B}_n}^0, \V{\beta}_{{\cal C}_n}^0 )}$ be
the underlying true parameters, where
\begin{eqnarray} \label{eq:crit13a}
 {\cal{A}}_n &=& \{ (k,j): \beta_{n,kj}^0 \neq 0 \}, \nonumber \\
 {\cal{B}}_n &=& \{ (k,j): \beta_{n,kj}^0 = 0, \V{\beta}_{nk}^0 \neq
 0 \}, \nonumber \\
 {\cal{C}}_n &=& \{ (k,j): \V{\beta}_{nk}^0 = 0 \}, \nonumber \\
 {\cal D}_n &=& {\cal B}_n \cup {\cal C}_n.
\end{eqnarray}
Note that ${\cal A}_n$ contains the indices of coefficients which
are truly non-zero, ${\cal C}_n$ contains the indices where the
whole ``groups'' are truly zero, and ${\cal B}_n$ contains the
indices of zero coefficients, but they belong to some non-zero
groups. So ${\cal A}_n$, ${\cal B}_n$ and ${\cal C}_n$ are disjoint
and they partition all the indices. We have the following theorem.
\begin{theorem} \label{thm2}
 If $ \sqrt{n} \lambda_n = O(1) $, then there exists a
 root-($n/P_n$) consistent local maximizer $\hat{\V{\beta}}_n =
 \T{(\hat{\V{\beta}}_{{\cal A}_n}, \hat{\V{\beta}}_{{\cal B}_n},
 \hat{\V{\beta}}_{{\cal C}_n})}$ of (\ref{asym:eq01n}), and if
 also $P_nn^{-3/4}/{\lambda_n} \rightarrow 0$ as $n \rightarrow \infty$,
 then $\F{Pr}(\hat{\V{\beta}}_{{\cal C}_n} = 0) \rightarrow 1$.
\end{theorem}
Theorem \ref{thm2} implies that the hierarchical lasso method can
effectively remove unimportant {\it groups}. For the above
root-($n/P_n$) consistent estimate, however, if ${{\cal B}_n} \neq
\emptyset$ (empty set), then $\F{Pr}(\hat{\V{\beta}}_{{\cal B}_n} =
0) \rightarrow 1$ is not always true. This means that although the
hierarchical lasso method can effectively remove {\it all}
unimportant {\it groups} and {\it some} of the unimportant {\it
variables} within important groups, it cannot effectively remove
{\it all} unimportant {\it variables} within important groups.

In the next section, we improve the hierarchical lasso method to
tackle this limitation.

\section{Adaptive Hierarchical Lasso}
\label{sec:improve}

To improve the hierarchical lasso method, we apply the adaptive idea
which has been used in \citeasnoun{Breiman95},
\citeasnoun{WangLiTsai06},
\citeasnoun{ZhangLu06},
and \citeasnoun{Zou06}, i.e., to penalize different coefficients
differently. Specifically, we consider
\begin{eqnarray} \label{eq:crit13b}
 \max_{\beta_{n,kj}} && - \frac{1}{2} \sum_{i=1}^{n} \left( y_{ni} -
 \sum_{k=1}^{K_n} \sum_{j=1}^{p_k} x_{ni,kj}\beta_{n,kj} \right)^2
 \nonumber \\
 && - n\lambda_n \cdot \sum_{k=1}^{K_n} \sqrt{ w_{n,k1} |\beta_{n,k1}| + w_{n,k2}
 |\beta_{n,k2}| + \ldots + w_{n,kp_k} |\beta_{n,kp_{nk}}| },
\end{eqnarray}
where $w_{n,kj}$ are pre-specified weights. The intuition is that if
the effect of a variable is strong, we would like the corresponding
weight to be small, hence the corresponding coefficient is lightly
penalized. If the effect of a variable is not strong, we would like
the corresponding weight to be large, hence the corresponding
coefficient is heavily penalized. In practice, we may consider using
the ordinary least squares estimates or the ridge regression
estimates to help us compute the weights, for example,
\begin{equation} \label{eq:wts}
 w_{n,kj} = \frac{1}{|\hat{\beta}_{n,kj}^{\textrm{ols}}|^\gamma}
 ~~~\textrm{or}~~~ w_{n,kj} =
 \frac{1}{|\hat{\beta}_{n,kj}^{\textrm{ridge}}|^\gamma},
\end{equation}
where $\gamma$ is a positive constant.

\subsection{Oracle Property}
\label{sec:theory2}

\subsubsection*{Problem Setup}
Since the theoretical results we develop for (\ref{eq:crit13b}) are
not restricted to the squared error loss, for the rest of the
section, we consider the generalized linear model. For generalized
linear models, statistical inferences are based on underlying
likelihood functions. We assume that the data
$\V{V}_{ni}=(\V{X}_{ni}, Y_{ni}), ~i=1, \ldots, n$ are independent
and identically distributed for every $n$. Conditioning on
$\V{X}_{ni} = \V{x}_{ni}$, $Y_{ni}$ has a density
$f_n(g_n(\T{\V{x}}_{ni} \V{\beta}_n), Y_{ni})$, where $g_n(\cdot)$
is a known link function. We maximize the penalized log-likelihood
\begin{eqnarray} \label{eq:crit131c}
 \max_{\beta_{n,kj}}~ Q_n(\V{\beta}_n) &=& L_n(\V{\beta}_n) - J_n(\V{\beta}_n)
 \nonumber \\
 &=& \sum_{i=1}^{n} \ell_n(g_n(\T{\V{x}}_{ni} \V{\beta}_n), y_{ni}) - n
 \sum_{k=1}^K p_{\lambda_n,\V{w}_n}(\V{\beta}_{nk} ),
\end{eqnarray}
where $\ell_n(\cdot,\cdot) = \log f_n(\cdot, \cdot)$ denotes the
conditional log-likelihood of $Y$, and
\[ p_{\lambda_n,\V{w}_n}(\V{\beta}_{nk}) = \lambda_n \sqrt{
 w_{n,k1} |\beta_{n,k1}| + \ldots + w_{n,kp_k} |\beta_{n,kp_{nk}}| }. \]

Note that under the normal distribution, $\ell_n(g_n(\T{\V{x}}_{ni}
\V{\beta}_n), y_{ni}) = - \frac{(y_{ni} - \T{\V{x}}_{ni}
\V{\beta}_n)^2}{2C_1} + C_2$, hence (\ref{eq:crit131c}) reduces to
(\ref{eq:crit13b}).

The asymptotic properties of (\ref{eq:crit131c}) are described in
the following theorems, and the proofs are in the Appendix. We note
that the proofs follow the spirit of \citeasnoun{FanLi01} and
\citeasnoun{FanPeng04}, but due to the grouping structure and the
adaptive weights, they are non-trivial extensions of
\citeasnoun{FanLi01} and \citeasnoun{FanPeng04}.

 To control the adaptive weights, we define:
\begin{eqnarray*}
 a_n &=& \max \{w_{n,kj}: \beta_{n,kj}^0 \neq 0 \}, \\
 b_n &=& \min \{w_{n,kj}: \beta_{n,kj}^0 = 0 \}.
\end{eqnarray*}

%
%
%

We assume that
$$ 0 < c_1 < \min \{ \beta_{n,kj}^0 : \beta_{n,kj}^0 \neq 0 \} <
 \max \{ \beta_{n,kj}^0 : \beta_{n,kj}^0 \neq 0 \} < c_2 <
 \infty. $$
Then we have the following results.

\begin{theorem} \label{thm_b1}
For every $n$, the observations  $\{ \V{V}_{ni}, i = 1, 2,
  \ldots, n \}$ are independent and identically distributed, each
with a density $f_n(\V{V}_{n1}, \V{\beta}_n)$
that satisfies conditions (A1)-(A3) in the Appendix. If $\frac{
{P^4_n} } {n} \rightarrow 0$ and $P_n^2 \lambda_n \sqrt{a_n} =
o_p(1)$, then there exists a local maximizer $\hat{\V{\beta}}_n$ of
$Q_n(\V{\beta}_n)$ such that $\| \hat{\V{\beta}}_n -\V{\beta}_n^0
\|=O_p( \sqrt {P_n} ( n^{-1/2}+ \lambda_n \sqrt{a_n}))$.
\end{theorem}

Hence by choosing $\lambda_n \sqrt{a_n} = O_p(n^{-1/2} ) $, there
exists a root-$(n/P_n)$ consistent penalized likelihood estimate.

\begin{theorem} \label{thm_b2}
For every $n$, the observations $\{ \V{V}_{ni}, i = 1, 2, \ldots, n \}$
are independent and identically
distributed, each with a density $f_n(\V{V}_{n1}, \V{\beta}_n)$
that satisfies conditions (A1)-(A3) in the Appendix. If $\frac{
{P^4_n} } {n} \rightarrow 0$, $\lambda_n \sqrt{a_n } = O_p(n^{-1/2}
)$ and $ \frac{ {P_n^{2}} }{\lambda_n^2 {b_n}} = o_p(n )$, then
there exists a root-$(n/P_n)$ consistent local maximizer
$\hat{\V{\beta}}_n$ such that:

\begin{enumerate}
\item[(a)] Sparsity: $\F{Pr}(\hat{\V{\beta}}_{n,{{\cal{D}}_n }} = 0)
  \rightarrow 1 $, where ${\cal D}_n = {\cal B}_n \cup {\cal C}_n$.
\item[(b)] Asymptotic normality: If $ \lambda_n \sqrt{a_n } =
  o_p( {(nP_n)}^{-1/2} )$ and $ \frac{P_n^5}{n}
  \rightarrow 0$ as $n\rightarrow \infty$, then we also have:
  \begin{equation*}
    \sqrt{n} \M{A}_n \M{I}_n^{1/2} ( {\V{\beta}}_{n,{{\cal{A}}_n }}^0 )
    ( \hat{\V{\beta}}_{n,{{\cal{A}}_n }} - {\V{\beta}}_{n,{{\cal{A}}_n }}^0 )
    \rightarrow {\cal{N}} (\V{0}, \M{G}),
  \end{equation*}
\end{enumerate}
where $\M{A}_n$ is a $q \times |{{\cal{A}}_n }| $ matrix such that
$\M{A}_n \T{\M{A}}_n \rightarrow \M{G}$ and $\M{G}$ is a $q\times
q$ nonnegative symmetric matrix. $\M{I}_n(
{\V{\beta}}_{n,{{\cal{A}}_n }}^0 )$ is the Fisher
information matrix knowing $\V{\beta}_{ {\cal{D}}_n }^0 = 0$.
\end{theorem}

The above requirements $\lambda_n \sqrt{a_n } = o_p( {(nP_n)}^{-1/2}
)$ and $\frac{ {P_n^{2}} }{\lambda_n^2 {b_n}} = o_p(n)$
as $n\rightarrow \infty$ can be satisfied by selecting $\lambda_n$
and ${w_{n,kj} }$ appropriately.  For example, we may let $\lambda_n
= \frac{(nP_n)^{-1/2}}{\mathrm{log}n} $ and $ {w_{n,kj} } =
\frac{1}{|\hat{\beta}^0_{n,kj}|^2}$, where $\hat{\beta}^0_{n,kj}$ is
the un-penalized likelihood estimate of ${\beta}_{n,kj}^0$, which is
root-($n/P_n$) consistent. Then we have $a_n = O_p(1) $ and
$\frac{1}{b_n} = O_p(\frac{P_n}{n})$. Hence $\lambda_n \sqrt{a_n } =
o_p({(nP_n)}^{-1/2})$ and $ \frac{
  {P_n^{2}} }{\lambda_n^2 {b_n}} = o_p(n )$ are satisfied when $ \frac{P_n^5}{n}
  \rightarrow 0$.

\subsection{Likelihood Ratio Test}

Similarly as in \citeasnoun{FanPeng04}, we develop a likelihood
ratio test for testing linear hypotheses:

$$ H_0:  \M{A}_n {\V{\beta}}_{n,{{\cal{A}}_n }}^0 = 0
\mathrm{~~~vs.~~~}
H_1: \M{A}_n {\V{\beta}}_{n,{{\cal{A}}_n }}^0 \ne 0 , $$
where $\M{A}_n$ is a $ q \times |{{\cal{A}}_n }| $ matrix and $
\M{A}_n \T{\M{A}}_n \rightarrow \M{I}_q$ for a fixed $q$. This
problem includes testing simultaneously the significance of
several covariate variables.

We introduce a natural likelihood ratio test statistic, i.e.
\[ T_n = 2 \left\{ \sup_{\Omega_n} Q_n(\V{\beta}_n|\V{V}) -
\sup_{\Omega_n, \M{A}_n {\V{\beta}}_{n,{{\cal{A}}_n }} = 0 }
Q_n(\V{\beta}_n|\V{V})  \right\} , \]
where $\Omega_n$ is the parameter space for $\V{\beta}_n$.  Then
we can obtain the following theorem regarding the asymptotic null
distribution of the test statistic.

\begin{theorem} \label{thm_b3}
When conditions in $(b)$ of Theorem \ref{thm_b2} are satisfied,
under $H_0$ we have
\[ T_n  \rightarrow \chi_q^2, ~~~\mathrm{as}~ n \rightarrow
\infty. \]
\end{theorem}

\section{Simulation Study} \label{sec:resultsimu}

In this section, we use simulations to demonstrate the hierarchical
lasso (HLasso) method, and compare the results with those of some
existing methods.

Specifically, we first compare hierarchical lasso with some other
group variable selection methods, i.e., the $L_2$-norm group lasso
(\ref{intro:eq05}) and the $L_\infty$-norm group lasso
(\ref{intro:eq07}).  Then we compare the adaptive hierarchical
lasso with some other ``oracle'' (but non-group variable
selection) methods, i.e., the SCAD and the adaptive lasso.

We extended the simulations in \citeasnoun{YuanLin06}.
We considered a model which had both categorical
and continuous prediction variables. We first generated seventeen independent
standard normal variables, $Z_1, \ldots, Z_{16}$ and W. The
covariates were then defined as $X_j=(Z_j+W)/\sqrt{2}$. Then the
last eight covariates $X_{9}, \ldots, X_{16}$ were discretized to
0, 1, 2, and 3 by $\Phi^{-1}(1/4)$, $\Phi^{-1}(1/2)$ and
$\Phi^{-1}(3/4)$. Each of $X_1,
\ldots, X_8$ was expanded through a fourth-order polynomial,
and only the main effects of $X_9, \ldots, X_{16}$ were
considered. This gave us a total of eight continuous groups with
four variables in each group and eight categorical groups with three
variables in each group. We considered two cases.
\begin{description}
\item[Case 1.] ``All-in-all-out''
  \begin{eqnarray*}
    Y &=& \left[ X_3 + 0.5X^2_3 + 0.1X^3_3 + 0.1X^4_3 \right] +
    \left[ X_6 - 0.5X^2_6 + 0.15X^3_6
      + 0.1X^4_6 \right] \\
    && + \left[ \mathbb{I}(X_{9}=0) + \mathbb{I}(X_{9}=1) +
      \mathbb{I}(X_{9}=2) \right] + \varepsilon.
  \end{eqnarray*}
\item[Case 2.] ``Not all-in-all-out''
  \begin{equation*}
    Y = \left( X_3 + X^2_3 \right) + \left( 2X_6 - 1.5X^2_6
    \right) + \left[ \mathbb{I}(X_{9}=0) +
      2~\mathbb{I}(X_{9}=1) \right] + \varepsilon.
  \end{equation*}
\end{description}

For all the simulations above, the error term $\varepsilon$ follows
a normal distribution $\F{N}(0, \sigma^2)$, where $\sigma^2$ was set
such that each of the signal to noise ratios, $\F{Var}(\T{\V{X}}
\V{\beta})/\F{Var}(\epsilon)$, was equal to 3. We generated $n=400$
training observations from each of the above models, along with 200
validation observations and 10,000 test observations. The validation
set was used to select the tuning parameters $\lambda$'s that
minimized the validation error. Using these selected $\lambda$'s, we
calculated the mean squared error (MSE) with the test set.
We repeated this 200 times and computed the average MSEs and their
corresponding standard errors. We also recorded how frequently the
important variables were selected and how frequently the unimportant
variables were removed. The results are summarized in Table
\ref{tab:3}.

As we can see, all shrinkage methods perform much better than OLS;
this illustrates that some regularization is crucial for prediction
accuracy. In terms of prediction accuracy, we can also see that when
variables in a group follow the ``all-in-all-out'' pattern, the
$L_2$-norm (group lasso) method performs slightly better than the
hierarchical lasso method (Case 1 of Table \ref{tab:3}). When
variables in a group do not follow the ``all-in-all-out'' pattern,
however, the hierarchical lasso method performs slightly better than
the $L_2$-norm method (Case 2 of Table \ref{tab:3}). For variable
selection, we can see that in terms of identifying important
variables, the four shrinkage methods, the lasso, the
$L_\infty$-norm, the $L_2$-norm, and the hierarchical lasso all
perform similarly (``Non-zero Var.'' of Table \ref{tab:3}). However,
the $L_2$-norm method and the hierarchical lasso method are more
effective at removing unimportant variables than lasso and the
$L_\infty$-norm method (``Zero Var.'' of Table \ref{tab:3}).

\begin{table}[tbh]
\caption{Comparison of several group variable selection methods,
  including the $L_2$-norm group lasso, the $L_\infty$-norm group
  lasso and the hierarchical lasso.  The OLS and the regular
  lasso are used as benchmarks.
  The upper part is for
  Case 1, and the lower part is for Case 2. ``MSE'' is the mean
  squared error on the test set. ``Zero Var.'' is the percentage
  of correctly removed unimportant variables. ``Non-zero Var.''
  is the percentage of correctly identified important variables.
  All the numbers outside parentheses are means over 200
  repetitions, and the numbers in the parentheses are the
  corresponding standard errors.}
\label{tab:3}
\begin{center}
\begin{tabular}{l|c|c|c|c|c}
\hline \hline
\multicolumn{6}{l}{Case 1: ``All-in-all-out''} \\
\hline & OLS & Lasso & $L_{\infty}$ & $L_2$ &HLasso
\\
\hline
MSE & 0.92 (0.018) & 0.47 (0.011) & 0.31 (0.008) & 0.18 (0.009) & 0.24 (0.008) \\
\hline
Zero Var. & - & 57\% (1.6\%) & 29\% (1.4\%) & 96\% (0.8\%) & 94\% (0.7\%) \\
\hline
Non-Zero Var. & - & 92\% (0.6\%) & 100\% (0\%) & 100\% (0\%) & 98\% (0.3\%) \\
\hline \hline
\multicolumn{6}{l}{Case 2: ``Not all-in-all-out''} \\
\hline
& OLS & Lasso & $L_{\infty}$ & $L_2$ &HLasso \\
\hline
MSE& 0.91 (0.018) & 0.26 (0.008) & 0.46 (0.012) & 0.21 (0.01) & 0.15 (0.006) \\
\hline
Zero Var.& - & 70\% (1\%) & 17\% (1.2\%) & 87\% (0.8\%) & 91\% (0.5\%) \\
\hline
Non-zero Var.& - & 99\% (0.3\%) & 100\% (0\%) & 100\% (0.2\%) & 100\% (0.1\%) \\
\hline \hline
\end{tabular}
\end{center}
\end{table}


In the above analysis, we used the criterion (\ref{model:eq02}) or
(\ref{asym:eq01}) for the hierarchical lasso, i.e., we did not use
the adaptive weights $w_{kj}$ to penalize different coefficients
differently. To assess the improved version of the hierarchical
lasso, i.e. criterion (\ref{eq:crit13b}), we also considered using
adaptive weights. Specifically, we applied the OLS weights in
(\ref{eq:wts}) to (\ref{eq:crit13b}) with $\gamma=1$.  We compared
the results with those of SCAD and the adaptive lasso, which also
enjoy the oracle property.  However, we note that SCAD and the
adaptive lasso do not take advantage of the grouping structure
information.  As a benchmark, we also computed the Oracle OLS
results, i.e., OLS using only the important variables. The results
are summarized in Table \ref{tab:4}.  We can see that in the
``all-in-all-out'' case, the adaptive hierarchical lasso removes
unimportant variables more effectively than SCAD and adaptive lasso,
and consequently, the adaptive hierarchical lasso outperforms SCAD
and adaptive lasso by a significant margin in terms of prediction
accuracy.  In the ``not all-in-all-out'' case, the advantage of
knowing the grouping structure information is reduced, however, the
adaptive hierarchical lasso still performs slightly better than SCAD
and adaptive lasso, especially in terms of removing unimportant
variables.

\begin{figure}[tbph]
\begin{center}
\psfig{file=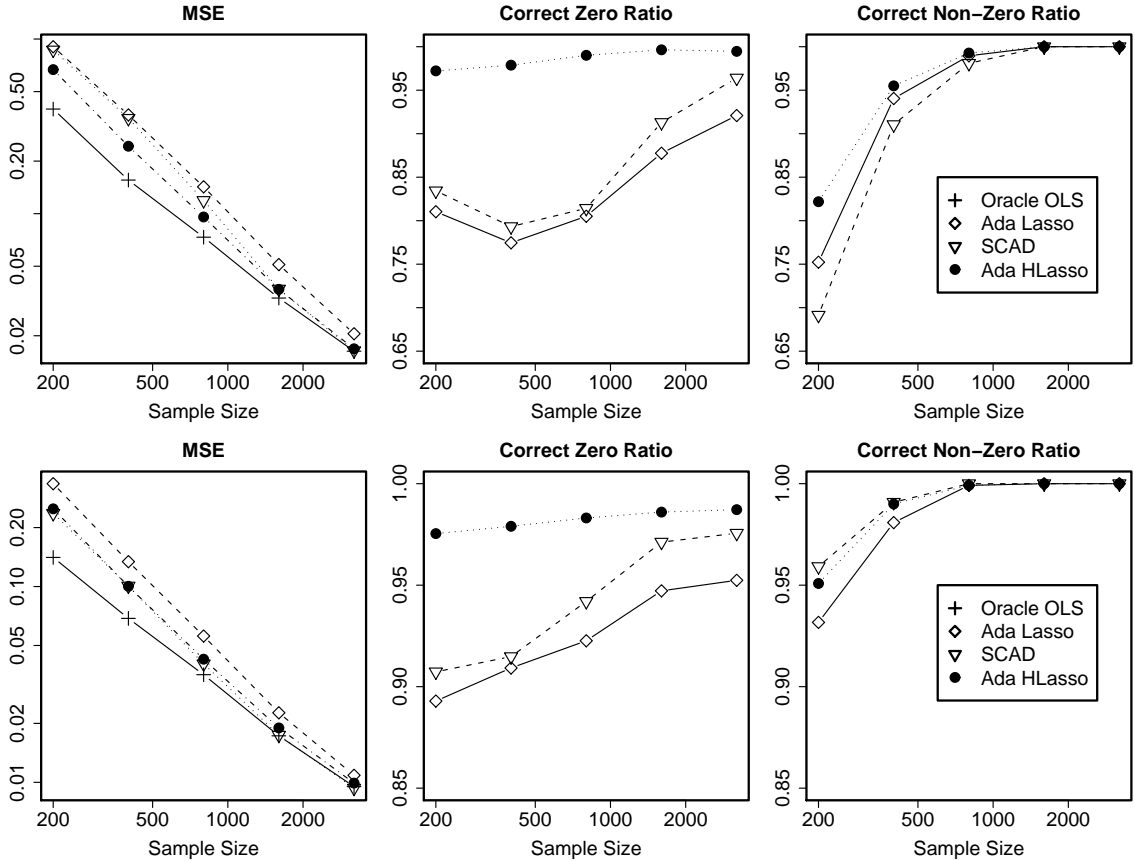, height=0.7\textheight, angle=270}
\caption{Comparison of several oracle methods, including the
  SCAD, the adaptive lasso and the adaptive hierarchical lasso.
  SCAD and adaptive lasso do not take advantage of the grouping
  structure information.
  The Oracle OLS uses only important variables.
  The first row corresponds to the ``all-in-all-out''
  case, and the second row corresponds to the
  ``not all-in-all-out'' case.
  ``Correct zero ratio'' records the percentage of correctly
  removed unimportant variables.  ``Correct non-zero ratio''
  records the percentage of correctly identified important
  variables.}
\label{fig:oracle}
\end{center}
\end{figure}

To assess how the sample size affects different ``oracle'' methods,
we also considered $n$=200, 400, 800, 1600 and 3200.  The results
are summarized in Figure \ref{fig:oracle}, where the first row
corresponds to the ``all-in-all-out'' case, and the second row
corresponds to the ``not all-in-all-out'' case.  Not surprisingly,
as the sample size increases, the performances of different methods
all improve: in terms of prediction accuracy, the MSE's all decrease
(at about the same rate) and get closer to that of the Oracle OLS;
in terms of variable selection, the probabilities of identifying the
correct model all increase and approach one. However, overall, the
adaptive hierarchical lasso always performs the best among the three
``oracle'' methods, and the gap is especially prominent in terms of
removing unimportant variables when the sample size is moderate.

\begin{table}[tbph]
\caption{Comparison of several ``oracle'' methods, including the
  adaptive hierarchical lasso, SCAD and the adaptive lasso.  SCAD
  and adaptive lasso do not take advantage of the grouping
  structure information.  The Oracle OLS uses only important
  variables.  Descriptions for the rows are the same as those in
  the caption of Table \ref{tab:3}.}
\label{tab:4}
\begin{center}
\begin{tabular}{l|c|c|c|c}
\hline \hline
\multicolumn{5}{l}{Case 1: ``All-in-all-out''} \\
\hline & Oracle OLS & Ada Lasso & SCAD & Ada HLasso \\
\hline
MSE & 0.16 (0.006) & 0.37 (0.011) & 0.35 (0.011) & 0.24 (0.009) \\
\hline
Zero Var. & - & 77\% (0.7\%) & 79\% (1.1\%) & 98\% (0.3\%) \\
\hline
Non-Zero Var. & - & 94\% (0.5\%) & 91\% (0.6\%) & 96\% (0.5\%) \\
\hline \hline
\multicolumn{5}{l}{Case 2: ``Not all-in-all-out''} \\
\hline
& Oracle OLS & Ada Lasso & SCAD & Ada HLasso \\
\hline
MSE& 0.07 (0.003) & 0.13 (0.005) & 0.11 (0.004) & 0.10 (0.005) \\
\hline
Zero Var.& - & 91\% (0.3\%) & 91\% (0.4\%) & 98\% (0.1\%) \\
\hline
Non-zero Var.& - & 98\% (0.4\%) & 99\% (0.3\%) & 99\% (0.3\%) \\
\hline \hline
\end{tabular}
\end{center}
\end{table}

\section{Real Data Analysis}\label{sec:resultreal}

In this section, we use a gene expression dataset from the NCI-60
collection of cancer cell lines to further illustrate the
hierarchical lasso method. We sought to use this dataset to identify
targets of the transcription factor p53, which regulates gene
expression in response to various signals of cellular stress. The
mutational status of the p53 gene has been reported for 50 of the
NCI-60 cell lines, with 17 being classified as normal and 33 as
carrying mutations \cite{OlivierEtal02}.

Instead of single-gene analysis, gene-set information has recently
been used to analyze gene expression data. For example,
\citeasnoun{SubramanianEtal05} developed the Gene Set Enrichment
Analysis (GSEA), which is found to be more stable and more powerful
than single-gene analysis. \citeasnoun{Efron07} improved the GSEA
method by using new statistics for summarizing gene-sets. Both
methods are based on hypothesis testing. In this analysis, we
consider using the hierarchical lasso method for gene-set selection.
The gene-sets used here are the cytogenetic gene-sets and the
functionals gene-sets from the GSEA package
\cite{SubramanianEtal05}. We considered 391 overlapping
gene-sets with the size of each set greater than 15.

Since the response here is binary (normal vs mutation), following
the result in Section \ref{sec:theory2}, we use the logistic hierarchical
lasso regression, instead of the least square hierarchical lasso.
Note that a gene may belong to multiple gene-sets, we thus extend
the hierarchical lasso to the case of overlapping groups. Suppose
there are $K$ groups and $J$ variables. Let $\mathcal{G}_k$ denote
the set of indices of the variables in the $k$th group. One way to
model the overlapping situation is to extend the criterion
(\ref{model:eq02}) as the following:
\begin{eqnarray}\label{logisticHlasso}
  \max_{d_k,\alpha_{j}} &&  \sum_{i=1}^{n} \ell
  \left(\sum_{k=1}^K d_k \sum_{j: j \in \mathcal{G}_k}
  \alpha_{j} x_{i,j},~y_{i}\right) \\ \nonumber
  && - \sum_{k=1}^K d_k - \lambda \cdot \sum_{j=1}^J |\alpha_j|
  \\ \nonumber
  \textrm{subject to} &&  d_k\ge 0, ~k=1, \ldots, K,
\end{eqnarray}
where $\alpha_j$ can be considered as the ``intrinsic'' effect of a
variable (no matter which group it belongs to), and different group
effects are represented via different $d_k$.  In this section,
$\ell(\eta_i,y_i) = y_i\eta_i - \log(1+e^{\eta_i})$ is the logistic
log-likelihood function with $y_i$ being a 0/1 response. Also notice
that if each variable belongs to only one group, the model reduces
to the non-overlapping criterion (\ref{model:eq02}).

We randomly split the 50 samples into the training and test sets 100
times; for each split, 33 samples (22 carrying mutations
and 11 being normal) were used for training and the rest 17 samples
(11 carrying mutations and 6 being normal) were for
testing. For each split, we applied three methods, the logistic
lasso, the logistic $L_2$-norm group lasso \cite{MeierEtAl08}
and the logistic hierarchical lasso.
Tuning parameters were chosen using five-fold cross-validation.

\begin{figure}[tbph]
\begin{center}
\psfig{file=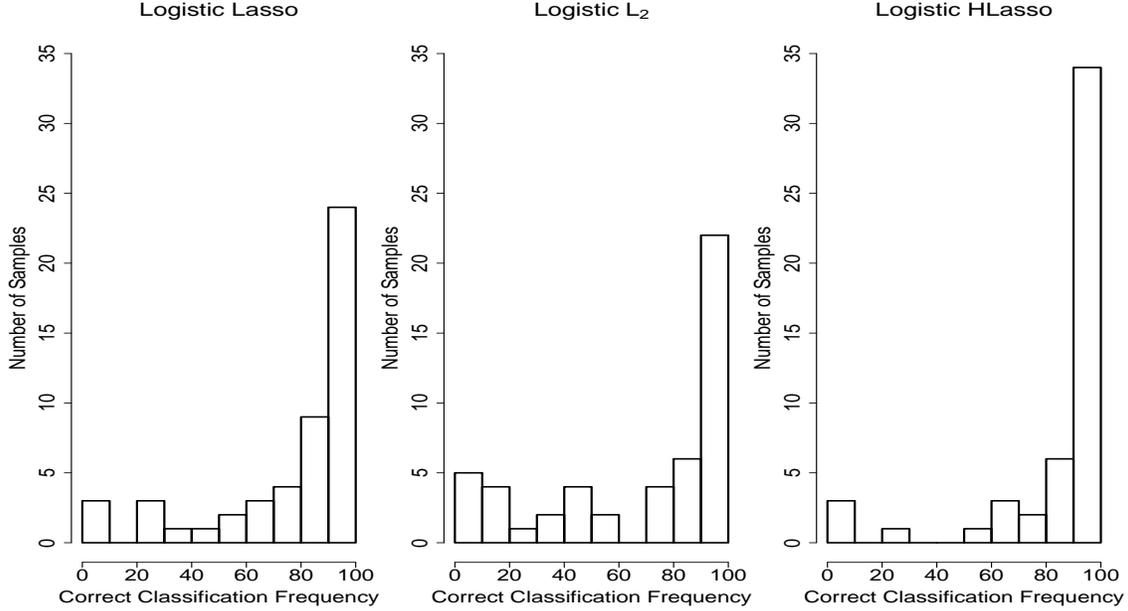,
  height=0.9\textwidth, width=0.5\textwidth, angle=270}
\caption{The number of samples vs the frequency that a
sample was correctly classified on 100 random splits of the p53
data. } \label{fig:freq}
\end{center}
\end{figure}

We first compare the prediction accuracy of the three methods. Over
the 100 random splits, the logistic hierarchical lasso has an
average misclassification rate of 14\% with the standard error
1.8\%, which is smaller than 23\%(1.7\%) of the logistic lasso and
32\%(1.2\%) of the logistic group lasso. To assess the stability of
the prediction, we recorded the frequency in which each sample, as a
test observation, was correctly classified. For example, if a sample
appeared in 40 test sets among the 100 random splits, and out of the
40 predictions, the sample was correctly classified 36 times, we
recorded 36/40 for this sample. The results are shown in Figure
\ref{fig:freq}. As we can see, for most samples, the logistic
hierarchical lasso classified them correctly for most of the random
splits, and the predictions seemed to be slightly more stable than
the logistic lasso and the logistic $L_2$-norm group lasso.

Next, we compare gene-set selection of these three methods.
The most notable difference is that both logistic lasso and the
logistic hierarchical lasso selected gene CDKN1A most frequently out
of the 100 random split, while the logistic $L_2$-norm group lasso
rarely selected it.  CDKN1A is also named as wild-type p53 activated
fragment-1 (p21), and it is known that the expression of gene CDKN1A
is tightly controlled by the tumor suppressor protein p53, through
which this protein mediates the p53-dependent cell cycle G1 phase
arrest in response to a variety of stress stimuli
\cite{LohEtAl03}.

We also compared the gene-sets selected by the logistic hierarchical
lasso with those selected by the GSEA of
\citeasnoun{SubramanianEtal05} and the GSA of \citeasnoun{Efron07}.
The two most frequently selected gene-sets by the hierarchical lasso
are \emph{atm pathway} and \emph{radiation sensitivity}. The most
frequently selected genes in \emph{atm pathway} by the logistic
hierarchical lasso are CDKN1A, MDM2 and RELA, and the most
frequently selected genes in \emph{radiation sensitivity} are
CDKN1A, MDM2 and BCL2.  It is known that MDM2, the second commonly
selected gene, is a target gene of the transcription factor tumor
protein p53, and the encoded protein in MDM2 is a nuclear
phosphoprotein that binds and inhibits transactivation by tumor
protein p53, as part of an autoregulatory negative feedback loop
\cite{KubbutatEtAl97,MollPetrenko03}.
Note that the gene-set \emph{radiation sensitivity} was also
selected by GSEA and GSA.  Though the gene-set \emph{atm pathway}
was not selected by GSEA and GSA, it shares 7, 8, 6, and 3 genes
with gene-sets \emph{radiation sensitivity}, \emph{p53 signalling},
\emph{p53 hypoxia pathway} and \emph{p53 Up} respectively, which
were all selected by GSEA and GSA. We also note that GSEA and GSA
are based on the {\it marginal} strength of each gene-set, while the
logistic hierarchical lasso fits an ``additive'' model and uses the
{\it joint} strengths of gene-sets.


\section{Discussion} \label{sec:summary}

In this paper, we have proposed a hierarchical lasso method for
group variable selection. Different variable selection methods have
their own advantages in different scenarios. The hierarchical lasso
method not only effectively removes unimportant groups, but also
keeps the flexibility of selecting variables within a group. We show
that the improved hierarchical lasso method enjoys an oracle
property, i.e., it performs as well as if the true sub-model were
given in advance. Numerical results indicate that our method works
well, especially when variables in a group are associated with the
response in a ``not all-in-all-out'' fashion.

The grouping idea is also applicable to other regression and
classification settings, for example, the multi-response regression
and multi-class classification problems. In these problems, a
grouping structure may not exist among the prediction variables, but
instead, natural grouping structures exist among {\it parameters}.
We use the multi-response regression problem to illustrate the point
\cite{BreimanFriedman97,TurlachVenablesWright05}. Suppose we observe
$(\V{x}_1, \V{y}_1)$, $\ldots$, $(\V{x}_n, \V{y}_n)$, where each
$\V{y}_i = (y_{i1}, \ldots, y_{iK})$ is a vector containing $K$
responses, and we are interested in selecting a subset of the
prediction variables that predict well for all of the multiple
responses. Standard techniques estimate $K$ prediction functions,
one for each of the $K$ responses, $f_k(\V{x}) = \beta_{k1} x_1 +
\cdots + \beta_{kp} x_p, k=1, \ldots, K$. The prediction variables
$(x_1, \ldots, x_p)$ may not have a grouping structure, however, we
may consider the coefficients corresponding to the same prediction
variable form a natural group, i.e., $(\beta_{1j}, \beta_{2j},
\ldots, \beta_{Kj})$. Using our hierarchical lasso idea, we
reparameterize $\beta_{kj} = d_j \alpha_{kj}$, $d_j \ge 0$, and we
consider
\begin{eqnarray*}
 \max_{d_j \ge 0, \alpha_{kj}} && - \frac{1}{2} \sum_{k=1}^K \sum_{i=1}^n
 \left( y_{ik} -
 \sum_{j=1}^p d_j \alpha_{kj} x_{ij} \right)^2 \\
 && - \lambda_1 \cdot \sum_{j=1}^p d_j
 - \lambda_2 \cdot \sum_{j=1}^p \sum_{k=1}^K |\alpha_{kj}|.
\end{eqnarray*}
Note that if $d_j$ is shrunk to zero, all $\beta_{kj}, k=1,\ldots,K$
will be equal to zero, hence the $j$th prediction variable will be
removed from all $K$ predictions. If $d_j$ is not equal to zero,
then some of the $\alpha_{kj}$ and hence some of the $\beta_{kj}$,
$k=1, \ldots, K$, still have the possibility of being zero.
Therefore, the $j$th variable may be predictive for some responses
but non-predictive for others.

One referee pointed out the work by \citeasnoun{HuangEtAl07}, which
we were not aware of when our manuscript was first completed and
submitted in 2007.  We acknowledge that the work by
\citeasnoun{HuangEtAl07} is closely related with ours, but there are
also differences.  For example:
\begin{itemize}
\item We proved the oracle property for both group selection
  and within group selection, while \citeasnoun{HuangEtAl07}
  considered the oracle property only for group selection.
\item Our theory applies to the generalized maximum likelihood
  estimate, while \citeasnoun{HuangEtAl07} considered the
  penalized least squares estimate.
\item Handling overlapping groups. It is
  not unusual for a variable to be a member of several
  groups. The gene expression date we analyzed in Section
  \ref{sec:resultreal} is such an example: given a plethora of
  biologically defined gene-sets, not surprisingly, there will be
  considerable overlap among these sets.

  In \citeasnoun{HuangEtAl07}, a prediction variable that appears
  in more than one group gets penalized more heavily than
  variables appearing in only one group.  Therefore, a prediction
  variable belonging to multiple groups is more likely to be
  removed than a variable belonging to only one group.  We are
  not sure whether this is an appealing property.
  In our approach, as shown in (\ref{logisticHlasso}),
  if a prediction variable
  belongs to multiple groups, it does not get penalized more
  heavily than other variables that belong to only one group.

\end{itemize}

\section*{Acknowledgments}
Zhou and Zhu were partially supported by grants DMS-0505432,
DMS-0705532 and DMS-0748389 from the National Science Foundation.

\bibliographystyle{agsm}

 \section*{Appendix}

\subsection*{Proof of Lemma 1}

Let $Q^*(\lambda_1, \lambda_2, \V{d}, \V{\alpha})$ be the criterion
that we would like to maximize in equation (7) and let
$Q^\star(\lambda, \V{d}, \V{\alpha})$ be the corresponding criterion
in equation (8).

Let ($\hat{\V{d}}^*, \hat{\V{\alpha}}^*$) be a local
 maximizer of
$Q^*(\lambda_1, \lambda_2, \V{d}, \V{\alpha})$. We would like to
prove ($\hat{\V{d}}^\star = \lambda_1\hat{\V{d}}^*,
\hat{\V{\alpha}}^\star = \hat{\V{\alpha}}^*/\lambda_1$) is a local
maximizer of $Q^\star(\lambda, \V{d}, \V{\alpha})$.

We immediately have $$Q^*(\lambda_1, \lambda_2, {\V{d}},
{\V{\alpha}}) = Q^\star(\lambda, \lambda_1{\V{d}},
 {\V{\alpha}}/\lambda_1).$$ Since ($\hat{\V{d}}^*, \hat{\V{\alpha}}^*$) is a local maximizer of
$Q^*(\lambda_1, \lambda_2, \V{d}, \V{\alpha})$, there exists $\delta
> 0$ such that if ${\V{d}}^{\prime}$, ${\V{\alpha}}^{\prime} $ satisfy
$ \| {\V{d}}^{\prime} - \hat{\V{d}}^* \| +
 \| {\V{\alpha}}^{\prime} - \hat{\V{\alpha}}^* \| < \delta $ then $
 Q^*(\lambda_1, \lambda_2, {\V{d}}^{\prime},
{\V{\alpha}}^{\prime} ) \le Q^*(\lambda_1, \lambda_2, \hat{\V{d}}^*,
\hat{\V{\alpha}}^*).$

Choose $\delta^{\prime} $ such that $
\frac{\delta^{\prime}}{\min\left( \lambda_1,
\frac{1}{\lambda_1}\right)} \le \delta$, for any
(${\V{d}}^{\prime\prime}, {\V{\alpha}}^{\prime\prime}$) satisfying $
\| {\V{d}}^{\prime\prime} - \hat{\V{d}}^\star \| + \|
{\V{\alpha}}^{\prime\prime} - \hat{\V{\alpha}}^\star \| <
\delta^{\prime} $ we have
\begin{eqnarray*}
\left\| \frac{ \V{d}^{\prime\prime} }{\lambda_1} -
 \hat{\V{d}}^* \right\| +
 \| { \lambda_1 \V{\alpha}}^{\prime\prime} -
 \hat{\V{\alpha}}^* \|
& \le & \frac{ \lambda_1 \left\| \frac{ \V{d}^{\prime\prime} }
 {\lambda_1} - \hat{\V{d}}^* \right\| +
 \frac{1}{\lambda_1} \left\| \lambda_1 \V{\alpha}^{\prime\prime} -
 \hat{\V{\alpha}}^* \right\| }
 { \min\left( \lambda_1, \frac{1}{\lambda_1}\right) }
 \\ 
& = & \frac{ \| {\V{d}}^{\prime\prime} - \hat{\V{d}}^\star \| +
 \| {\V{\alpha}}^{\prime\prime} - \hat{\V{\alpha}}^\star \| }
 { \min\left( \lambda_1, \frac{1}{\lambda_1}\right) }
 \\ 
& < & \frac{ \delta^{\prime}} { \min\left( \lambda_1,
 \frac{1}{\lambda_1}\right) }
 \\ 
& < & \delta.
\end{eqnarray*}
Hence \begin{eqnarray*}
 Q^\star(\lambda, \hat{\V{d}}^{\prime\prime}, \hat{\V{\alpha}}^{\prime\prime})
 &=& Q^\ast (\lambda_1, \lambda_2, \hat{\V{d}}^{\prime\prime}/\lambda_1,
 \lambda_1\hat{\V{\alpha}}^{\prime\prime}) \\
 &\le & Q^\ast (\lambda_1, \lambda_2, \hat{\V{d}}^\ast,
 \hat{\V{\alpha}}^\ast) \\
 &=&
 Q^\star(\lambda, \hat{\V{d}}^\star, \hat{\V{\alpha}}^\star).
\end{eqnarray*} Therefore, ($\hat{\V{d}}^\star = \lambda_1\hat{\V{d}}^*,
\hat{\V{\alpha}}^\star = \hat{\V{\alpha}}^*/\lambda_1$) is a local
maximizer of $Q^\star(\lambda, \V{d}, \V{\alpha}).$

Similarly we can prove that for any local maximizer
($\hat{\V{d}}^\star, \hat{\V{\alpha}}^\star$) of $Q^\star(\lambda,
\V{d}, \V{\alpha})$, there is a corresponding local maximizer
($\hat{\V{d}}^*, \hat{\V{\alpha}}^*$) of $Q^*(\lambda_1, \lambda_2,
\V{d}, \V{\alpha})$ such that $ \hat{d}_k^{\ast}
\hat{\alpha}_{kj}^{\ast} = \hat{d}_k^\star
 \hat{\alpha}_{kj}^\star. $

\begin{lemma} \label{lemma1}
Suppose ($\hat{\V{d}}, \hat{\V{\alpha}} $) is a local maximizer of
(8). Let $\hat{\V{\beta}} $ be the Hierarchical Lasso estimate
related to ($\hat{\V{d}}, \hat{\V{\alpha}} $), i.e.,
$\hat{\beta}_{kj} = \hat{d}_k \hat{\alpha}_{kj} $. If $\hat{d}_k =
0$, then $\hat{\V{\alpha}}_k =0 $; if $\hat{d}_k \ne 0$, then $\|
\hat{\V{\beta}}_{k} \|_1 \ne 0$ and $\hat{d}_k = \sqrt{ \lambda \|
\hat{\V{\beta}}_{k} \|_1 }, \hat{\V{\alpha}}_k = \frac{
\hat{\V{\beta}}_k }{ \sqrt{\lambda \| \hat{\V{\beta}}_{k} \|_1 } }
$.
\end{lemma}

\subsection*{Proof of Lemma 2}

If $\hat{d}_k = 0$, then $\hat{\V{\alpha}}_k =0 $ is quite obvious.
Similarly, if $\hat{\V{\alpha}}_k =0 $, then $\hat{d}_k = 0$.
Therefore, if $\hat{d}_k \ne 0$, then $\hat{\V{\alpha}}_k \ne 0 $
and $\| \hat{\V{\beta}}_{k} \|_1 \ne 0$.

We prove $\hat{d}_k = \sqrt{ \lambda \| \hat{\V{\beta}}_{k} \|_1 },
\hat{\V{\alpha}}_k = \frac{ \hat{\V{\beta}}_k }{ \sqrt{\lambda \|
\hat{\V{\beta}}_{k} \|_1 } } $ for $\hat{d}_k \ne 0$ by
contradiction. Suppose $\exists k^{\prime}$ such that
$\hat{d}_{k^\prime} \ne 0 $ and $\hat{d}_{k^\prime} \ne \sqrt{
\lambda \| \hat{\V{\beta}}_{k^\prime} \|_1 }$. Let $ \frac{ \sqrt{
\lambda \| \hat{\V{\beta}}_{k^\prime} \|_1 } }{ \hat{d}_{k^\prime} }
= c $. Then $\hat{\V{\alpha}}_k = c \frac{ \hat{\V{\beta}}_k }{
\sqrt{\lambda \| \hat{\V{\beta}}_{k} \|_1 } } $. Suppose $c > 1$.

Let $ \tilde{d}_{k} = \hat{d}_{k} $ and $ \tilde{\V{\alpha}}_{k} =
\hat{\V{\alpha}}_{k} $ for $k\ne k^\prime$ and $
\tilde{d}_{k^\prime} = \delta^\prime \hat{d}_{k^\prime} $ and $
\tilde{\V{\alpha}}_{k^\prime} = \hat{\V{\alpha}}_{k^\prime} \frac{ 1
}{ \delta^\prime }$, where $\delta^\prime$ satisfies $ c >
\delta^\prime > 1 $ and is very close to 1 such that $ \|
\tilde{d}_{k^\prime} - \hat{d}_{k^\prime} \|_1 + \|
\tilde{\V{\alpha}}_{k^\prime} - \hat{\V{\alpha}}_{k^\prime} \|_1 <
\delta $ for some $\delta >0 $.

Then we have
\begin{eqnarray*}
 Q^\star(\lambda, \tilde{\V{d}}, \tilde{\V{\alpha}}) -
 Q^\star(\lambda, \hat{\V{d}}, \hat{\V{\alpha}})
 &=& - \delta^\prime | \hat{d}_{k^\prime} | -
 \frac{1}{\delta^\prime} \lambda \| \hat{\V{\alpha}}_{k^\prime} \|_1 +
 | \hat{d}_{k^\prime} | +
 \lambda \| \hat{\V{\alpha}}_{k^\prime} \|_1 \\
 &=& \left( - \frac{\delta^\prime }{ c } - \frac{c}{\delta^\prime} +
 \frac{ 1 }{ c } + c \right) \sqrt{ \lambda \| \hat{\V{\beta}}_{k^\prime}\|_1 } \\
 &=& \frac{ 1 }{ c }( \delta^\prime - 1 )
 \left(\frac{c^2}{\delta^\prime} - 1 \right) \sqrt{ \lambda \|
 \hat{\V{\beta}}_{k^\prime} \|_1 } \\
 & > & 0.
\end{eqnarray*}
Therefore, for any $\delta > 0 $, we can find
$\tilde{\V{d}},\tilde{\V{\alpha}}$ such that $ \| \tilde{\V{d}} -
\hat{\V{d}} \|_1 + \|\tilde{\V{\alpha}} - \hat{\V{\alpha}} \|_1 <
\delta $ and $Q^\star(\lambda, \tilde{\V{d}}, \tilde{\V{\alpha}}) >
Q^\star(\lambda, \hat{\V{d}}, \hat{\V{\alpha}}) $. These contradict
with $(\hat{\V{d}},\hat{\V{\alpha}})$ being a local maximizer.

Similarly for the case when $ c < 1 $. Hence, we have the result
that if $\hat{d}_k \ne 0$, then $\hat{d}_k = \sqrt{ \lambda \|
\hat{\V{\beta}}_{k} \|_1 }, \hat{\V{\alpha}}_k = \frac{
\hat{\V{\beta}}_k }{ \sqrt{\lambda \| \hat{\V{\beta}}_{k} \|_1 } }
$.

\subsection*{Proof of Theorem 1}

Let $ Q(\lambda, \V{\beta})$ be the corresponding criterion in
equation (11).

Suppose ($\hat{\V{d}}, \hat{\V{\alpha}}$) is a local maximizer of
$Q^\star(\lambda, \V{d}, \V{\alpha})$, we first show that
$\hat{\V{\beta}}$, where $\hat{\beta}_{kj}=
\hat{d}_k\hat{\alpha}_{kj} $, is a local maximizer of $
Q(\lambda,\V{\beta})$, i.e. there exists a $ \delta^\prime $ such
that if $ \| \triangle\V{\beta} \|_1 < \delta^\prime $ then
$Q(\lambda, \hat{\V{\beta}} + \triangle\V{\beta} ) \le
Q(\lambda,\hat{\V{\beta}})$.

We denote $\triangle\V{\beta} = \triangle\V{\beta}^{(1)} +
\triangle\V{\beta}^{(2)} $, where $\triangle\V{\beta}^{(1)}_k = 0$
if $\| \hat{\V{\beta}}_{k} \|_1 = 0 $ and
$\triangle\V{\beta}^{(2)}_k = 0$ if $\| \hat{\V{\beta}}_{k} \|_1 \ne
0 $. We have $ \| \triangle\V{\beta} \|_1 = \|
\triangle\V{\beta}^{(1)} \|_1 + \| \triangle\V{\beta}^{(2)} \|_1 $.

Now we show $Q(\lambda, \hat{\V{\beta}} + \triangle\V{\beta}^{(1)} )
\le Q(\lambda, \hat{\V{\beta}})$ if $ \delta^\prime $ is small
enough. By Lemma 2, we have $\hat{d}_k = \sqrt{ \lambda \|
\hat{\V{\beta}}_{k} \|_1 }, \hat{\V{\alpha}}_k =
\frac{\hat{\V{\beta}}_k } { \sqrt{\lambda \| \hat{\V{\beta}}_{k}
\|_1} } $ if $\| \hat{d}_{k} \|_1 \ne 0 $ and $\hat{\V{\alpha}}_k =
\V{0} $ if $\| \hat{d}_{k} \|_1 = 0$. Furthermore, let
$\hat{d}_k^\prime =
 \sqrt{ \lambda \| \hat{\V{\beta}}_{k} + \triangle\V{\beta}^{(1)}_k \|_1 },
\hat{\V{\alpha}}_k^\prime = \frac{ \hat{\V{\beta}}_k +
\triangle\V{\beta}^{(1)}_k } { \sqrt{\lambda \| \hat{\V{\beta}}_{k}
+ \triangle\V{\beta}^{(1)}_k \|_1 } } $ if $\| \hat{d}_{k} \|_1 \ne
0 $. Let $\hat{d}_k^\prime = 0, \hat{\V{\alpha}}_k^\prime = \V{0} $
if $\| \hat{d}_{k} \|_1 = 0$. Then we have $Q^\star(\lambda,
\hat{\V{d}}^\prime, \hat{\V{\alpha}}^\prime ) = Q(\lambda,
\hat{\V{\beta}} + \triangle\V{\beta}^{(1)} )$ and $ Q^\star(\lambda,
\hat{\V{d}}, \hat{\V{\alpha}}) = Q(\lambda, \hat{\V{\beta}}) $.
Hence we only need to show that $Q^\star(\lambda,
\hat{\V{d}}^\prime, \hat{\V{\alpha}}^\prime ) \le Q^\star(\lambda,
\hat{\V{d}}, \hat{\V{\alpha}}) $. Note that ($\hat{\V{d}},
\hat{\V{\alpha}}$) ia a local maximizer of $Q^\star(\lambda,
\V{d},\V{\alpha})$. Therefore there exists a $\delta$ such that for
any ${\V{d}}^{\prime}, {\V{\alpha}}^{\prime}$ satisfying
$\|{\V{d}}^{\prime} - \hat{\V{d}} \|_1 +
 \| {\V{\alpha}}^{\prime} - \hat{\V{\alpha}} \|_1 <
 \delta $, we have $Q^\star(\lambda, {\V{d}}^\prime, {\V{\alpha}}^\prime ) \le
Q^\star(\lambda, \hat{\V{d}}, \hat{\V{\alpha}}) $.

Now since
\begin{eqnarray*}
 |\hat{d}_k^\prime - \hat{d}_k| &=& | \sqrt{ \lambda \|
\hat{\V{\beta}}_{k} + \triangle\V{\beta}^{(1)}_k \|_1 } - \sqrt{
\lambda \| \hat{\V{\beta }}_{k}
\|_1 }| \\
 & \le & | \sqrt{ \lambda \|
\hat{\V{\beta}}_{k} \|_1 - \lambda \| \triangle\V{\beta}^{(1)}_k
\|_1 } - \sqrt{ \lambda \| \hat{\V{\beta }}_{k}
\|_1 }| \\
 & \le & \frac{1}{2} \frac{ \lambda \|
\triangle\V{\beta}^{(1)}_k \|_1 }{ \sqrt{ \lambda \|
\hat{\V{\beta}}_{k} \|_1 - \lambda \| \triangle\V{\beta}^{(1)}_k
\|_1 }} \\
 & \le & \frac{1}{2} \frac{ \lambda \|
\triangle\V{\beta}^{(1)}_k
\|_1 }{ \sqrt{ \lambda a - \lambda \delta^\prime }} \\
 & \le & \frac{1}{2} \frac{ \lambda \|
\triangle\V{\beta}^{(1)}_k \|_1 }{ \sqrt{ \lambda a/2 }},
\end{eqnarray*}
where $ a = \min \{ \| \hat{\V{\beta}}_{k} \|_1 : \|
\hat{\V{\beta}}_{k} \|_1 \ne 0 \} $ and $ \delta^\prime < a/2$.

Furthermore
\begin{eqnarray*}
 \| \hat{\V{\alpha}}_k^\prime - \hat{\V{\alpha}}_k \|_1 &=&
 \left\| \frac{ \hat{\V{\beta}}_k + \triangle\V{\beta}^{(1)}_k }{
\sqrt{\lambda \| \hat{\V{\beta}}_{k} + \triangle\V{\beta}^{(1)}_k
\|_1 } } - \frac{ \hat{\V{\beta}}_k }{ \sqrt{\lambda \|
\hat{\V{\beta}}_{k}
\|_1 } } \right\|_1 \\ 
 & \le & \left\| \frac{ \hat{\V{\beta}}_k + \triangle\V{\beta}^{(1)}_k }{
\sqrt{\lambda \| \hat{\V{\beta}}_{k} + \triangle\V{\beta}^{(1)}_k
\|_1 } } - \frac{ \hat{\V{\beta}}_k }{ \sqrt{\lambda \|
\hat{\V{\beta}}_{k} + \triangle\V{\beta}^{(1)}_k
\|_1 } } \right\|_1 \\ 
 & & + \left\| \frac{ \hat{\V{\beta}}_k }{
\sqrt{\lambda \| \hat{\V{\beta}}_{k} + \triangle\V{\beta}^{(1)}_k
\|_1 } } - \frac{ \hat{\V{\beta}}_k }{ \sqrt{\lambda \|
\hat{\V{\beta}}_{k}
\|_1 } } \right\|_1 \\ 
 & \le & \frac{ \|
\triangle\V{\beta}^{(1)}_k
\|_1 }{ \sqrt{ \lambda a/2 }} \\ 
 & & + \frac{ \| \hat{\V{\beta}}_k \|_1 | \sqrt{ \lambda \|
\hat{\V{\beta}}_{k} + \triangle\V{\beta}^{(1)}_k \|_1 } - \sqrt{
\lambda \| \hat{\V{\beta }}_{k} \|_1 }| }{ \sqrt{ \lambda \|
\hat{\V{\beta}}_{k} + \triangle\V{\beta}^{(1)}_k \|_1 }
\sqrt{\lambda \| \hat{\V{\beta}}_{k}
\|_1 } } \\ 
 & \le & \frac{ \|
\triangle\V{\beta}^{(1)}_k \|_1 }{ \sqrt{ \lambda a/2 }} + \frac{ b
}{ \sqrt{ \lambda a/2 } \sqrt{ \lambda a} } \left( \frac{1}{2}
\frac{ \lambda \| \triangle\V{\beta}^{(1)}_k
\|_1 }{ \sqrt{ \lambda a/2 }} \right) \\ 
 & \le & \|
\triangle\V{\beta}^{(1)}_k \|_1 \left( \frac{ 1 }{ \sqrt{ \lambda
a/2 }} + \frac{ b }{ a \sqrt{ \lambda a } } \right),
\end{eqnarray*}
where $ b = \max \{ \| \hat{\V{\beta}}_{k} \|_1 : \|
\hat{\V{\beta}}_{k} \|_1 \ne 0 \} $.

Therefore, there exists a small enough $\delta^\prime$, if $ \|
\triangle\V{\beta}^{(1)} \|_1 < \delta^\prime $ we have $\|
\hat{\V{d}}^{\prime} - \hat{\V{d}} \|_1 +
 \| \hat{\V{\alpha}}^{\prime} - \hat{\V{\alpha}} \|_1 < \delta $.
Hence $ Q^\star(\lambda, \hat{\V{d}}^\prime, \hat{\V{\alpha}}^\prime
) \le Q^\star(\lambda, \hat{\V{d}}, \hat{\V{\alpha}}) $ (due to
local maximality) and $Q(\lambda, \hat{\V{\beta}} +
\triangle\V{\beta}^{(1)} ) \le Q(\lambda, \hat{\V{\beta}})$.

Next we show $ Q(\lambda, \hat{\V{\beta}} + \triangle\V{\beta}^{(1)}
+ \triangle\V{\beta}^{(2)} ) \le Q(\lambda, \hat{\V{\beta}} +
\triangle\V{\beta}^{(1)} ).$ Note that
\[
 Q(\lambda, \hat{\V{\beta}} + \triangle\V{\beta}^{(1)} +
 \triangle\V{\beta}^{(2)} ) -
Q(\lambda, \hat{\V{\beta}} + \triangle\V{\beta}^{(1)} ) = \T{
\triangle\V{\beta}^{(2)} } \nabla L( \hat{\V{\beta}}^* ) -
\sum_{k=1}^{K} \sqrt{ \lambda \| \triangle\V{\beta}^{(2)} \|_1 }, \]
where $\V{\beta}^*$ is a vector between $\hat{\V{\beta}} +
\triangle\V{\beta}^{(1)} + \triangle\V{\beta}^{(2)}$ and
$\hat{\V{\beta}} + \triangle\V{\beta}^{(1)} $. Since $ \|
\triangle\V{\beta}^{(2)} \|_1 < \delta^\prime $ is small enough, the
second term dominates the first term, hence we have $ Q(\lambda,
\hat{\V{\beta}} + \triangle\V{\beta}^{(1)} +
\triangle\V{\beta}^{(2)} ) \le Q(\lambda, \hat{\V{\beta}} +
\triangle\V{\beta}^{(1)} ) $.

Overall, we have that there exists a small enough $\delta^\prime$,
if $ \| \triangle\V{\beta} \|_1 < \delta^\prime $, then $ Q(\lambda,
\hat{\V{\beta}} + \triangle\V{\beta} ) \le Q(\lambda,
\hat{\V{\beta}} ) $, which implies that $\hat{\V{\beta}}$ is a local
maximizer of $Q(\lambda, {\V{\beta}})$.

Similarly, we can prove that if $\hat{\V{\beta}}$ is a local
maximizer of $Q(\lambda, {\V{\beta}})$, and if we let
 $\hat{d}_k = \sqrt{ \lambda \| \hat{\V{\beta}}_{k} \|_1 },
 \hat{\V{\alpha}}_k = \frac{\hat{\V{\beta}}_k }
 { \sqrt{\lambda \| \hat{\V{\beta}}_{k} \|_1} } $
for $\| \hat{\V{\beta}}_{k} \|_1 \ne 0 $ and let $\hat{d}_k = 0,
\hat{\V{\alpha}}_k = \V{0} $ for $\| \hat{\V{\beta}}_{k} \|_1 = 0$,
then ($\hat{\V{d}}, \hat{\V{\alpha}}$) is a local maximizer of
$Q^\star(\lambda, \V{d}, \V{\alpha})$.

\subsection*{Regularity Conditions }

Let $S_n$ be the number of non-zero groups, i.e., $\|
\V{\beta}_{nk}^0 \| \neq 0$. Without loss of generality, we assume
\begin{eqnarray*}
 \| \V{\beta}_{nk}^0 \| &\neq& 0, ~\textrm{for}~ k=1, \ldots, S_n, \\
 \| \V{\beta}_{nk}^0 \| &=& 0, ~\textrm{for}~ k=S_n+1, \ldots, K_n.
\end{eqnarray*}
Let $s_{nk}$ be the number of non-zero coefficients in group $k,
1\le k \le S_n$; again, without loss of generality, we assume
\begin{eqnarray*}
 \beta_{n,kj}^0 &\neq& 0, ~\textrm{for}~ k=1, \ldots, S_n;~ j = 1,
 \ldots, s_{nk}, \\
 \beta_{n,kj}^0 &=& 0, ~\textrm{for}~ k=1, \ldots, S_n;~ j=s_{nk}+1, \ldots, p_{nk}.
\end{eqnarray*}

 For simplicity, we write
$\beta_{n,kj}$, $p_{nk}$ and $s_{nk}$ as $\beta_{ kj}$, $p_k$ and
$s_k$ in the following.

Since we have diverging number of parameters, to keep the uniform
properties of the likelihood function, we need some conditions on
the higher-order moment of the likelihood function, as compared to
the usual condition in the asymptotic theory of the likelihood
estimate under finite parameters (Lehmann and Casella 1998).

\begin{itemize}
 \item[(A1)] For every $n$, the observations  $\{ \V{V}_{ni}, i = 1, 2,
  \ldots, n \}$ are independent and identically distributed, each
with a density $f_n(\V{V}_{n1}, \V{\beta}_n)$. $f_n(\V{V}_{n1},
\V{\beta}_n)$
 has a common support and the model is identifiable. Furthermore, the first and second
 logarithmic derivatives of $f_n$ satisfy the equations
 \begin{eqnarray*}
 \F{E}_{{\V{\beta}_n}}\left[\frac{\partial \log f_n(\V{V}_{n1},
\V{\beta}_n) } {\partial\beta_{kj}}\right] &=& 0,
 ~~~\textrm{for}~ k=1, \ldots, {K_n};~ j=1, \ldots, p_k \\
 \M{I}_{k_1j_1k_2j_2} (\V{\beta}_n) &=& \F{E}_{\V{\beta}_n} \left[
 \frac{\partial}{\partial\beta_{k_1j_1}} \log f_n(\V{V}_{n1}, \V{\beta}_n)
 \frac{\partial}{\partial\beta_{k_2j_2}} \log f_n(\V{V}_{n1},
 \V{\beta}_n) \right] \\
 &=& \F{E}_{\V{\beta}_n} \left[ -
 \frac{\partial^2}{\partial\beta_{k_1j_2} \partial\beta_{k_2j_2}}
 \log f_n(\V{V}_{n1}, \V{\beta}_n) \right].
 \end{eqnarray*}

 \item[(A2)] The Fisher information matrix
\begin{equation}
 \M{I}(\V{\beta}_n) =
\F{E}_{{\V{\beta}_n}}\left[ \frac{\partial}{\partial{\V{\beta}_n}}
\log f_n(\V{V}_{n1}, \V{\beta}_n)
 \frac{ \T{\partial} }{\partial{\V{\beta}_n}} \log
f_n(\V{V}_{n1}, \V{\beta}_n) \right]
 \\ \nonumber
\end{equation}
satisfies the condition
$$ 0 < C_1 < \lambda_{\min} \{ \M{I}(\V{\beta}_n) \} \le
\lambda_{\max} \{ \M{I}(\V{\beta}_n) \} < C_2 < \infty ,$$ and for
any $k_{1}, j_{1}, k_{2}, j_{2}$, we have
\begin{eqnarray*}
 \F{E}_{{\V{\beta}_n}}\left[\frac{\partial}{\partial\beta_{k_{1}j_{1}}}
\log f_n(\V{V}_{n1}, \V{\beta}_n) \frac{\partial}
{\partial\beta_{k_{2}j_{2}}} \log f_n(\V{V}_{n1}, \V{\beta}_n)
\right]^2 & < & C_3 < \infty,
\\ 
 \F{E}_{{\V{\beta}_n}}\left[- \frac{\partial^2}
 {\partial\beta_{k_{1}j_{1}}\partial\beta_{k_{2}j_{2}}}
\log f_n(\V{V}_{n1}, \V{\beta}_n) \right]^2 & < & C_4 < \infty.
 \end{eqnarray*}

 \item[(A3)] There exists an open subset $\omega_n$ of $\Omega_n \in R^{P_n}$
 that contains the true parameter
 point $\V{\beta}_n^0$ such that for almost all $\V{V}_{n1}$,
 the density $f_n(\V{V}_{n1}, \V{\beta}_n)$
 admits all third derivatives $\partial^3 f_n(\V{V}_{n1}, \V{\beta}_n)/
(\partial\beta_{k_{1}j_{1}}\partial\beta_{k_{2}j_{2}}\partial\beta_{k_{3}j_{3}})$
for all ${\V{\beta}_n}\in\omega_n $. Furthermore, there exist
functions $M_{{nk_{1}j_{1}}{k_{2}j_{2}}{k_{3}j_{3}}}$ such that
\begin{equation}
 \left|\frac{\partial^3}{\partial\beta_{k_{1}j_{1}}\partial\beta_{k_{2}j_{2}} \partial\beta_{k_{3}j_{3}}}
\log f_n(\V{V}_{n1}, \V{\beta}_n) \right| \leq
M_{{nk_{1}j_{1}}{k_{2}j_{2}}{k_{3}j_{3}}}(\V{V}_{n1})
 ~~ \mathrm{for~ all~} {\V{\beta}_n}\in\omega_n,
 \\ \nonumber
\end{equation}
and $ \F{E}_{{\V{\beta}_n}}[
M_{{nk_{1}j_{1}}{k_{2}j_{2}}{k_{3}j_{3}}}^2(\V{V}_{n1}) ]
 < C_5 < \infty $.

\end{itemize}

These regularity conditions guarantee the asymptotic normality of
the ordinary maximum likelihood estimates for diverging number of
parameters.

\bigskip

For expositional simplicity, we will first prove Theorem 3 and
Theorem 4, then prove Theorem 2.

\subsection*{Proof of Theorem 3}
%

We will show that for any given $\epsilon > 0$, there exists a
constant $C$ such that
\begin{equation} \label{sec_app_infty072}
\F{Pr}\left\{ \sup_{ \| \V{u}\| = C} Q_n(\V{\beta}_n^0 + \alpha_n
\V{u} ) < Q_n(\V{\beta}_n^0) \right\} \ge 1- \epsilon,
\end{equation}
where $\alpha_n = \sqrt{P_n} ( n^{-1/2} + \lambda_n \sqrt{a_n}
 /2\sqrt{c_1} ) $.
This implies that with probability at least $1- \epsilon $, that
there exists a local maximum in the ball $\{ \V{\beta}_n^0 +
\alpha_n \V{u} : {\|\V{u}\| \le C} \} $.
 Hence, there exists a local maximizer such that $\| \hat{\V{\beta}}_n -\V{\beta}_n^0
 \| = O_p(\alpha_n).$
Since $1/2\sqrt{c_1}$ is a constant, we have $\| \hat{\V{\beta}}_n
-\V{\beta}_n^0
 \| = O_p( \sqrt{P_n} (n^{-1/2} + \lambda_n \sqrt{a_n} ) ) $.


Using $p_{\lambda_n,\V{w}_n}(0) = 0 $, we have
\begin{eqnarray} \label{sec_app_infty0072a}
 D_n(\V{u}) &=& Q_n(\V{\beta}_n^0 + \alpha_n \V{u} ) - Q_n(\V{\beta}_n^0)
 \nonumber \\
 &\le& L_n(\V{\beta}_n^0 + \alpha_n \V{u} ) - L_n(\V{\beta}_n^0) \nonumber \\
 & & ~~~ - n \sum_{k=1}^{S_n}
 ( p_{\lambda_n,\V{w}_n}( {\V{\beta}}_{nk}^0+ \alpha_n \V{u}_k )
 - p_{\lambda_n,\V{w}_n}({\V{\beta}}_{nk}^0 ) ) \nonumber \\
 & \triangleq & (I) + (II).
\end{eqnarray}

 Using the standard argument on the Taylor
expansion of the likelihood function, we have
\begin{eqnarray} \label{sec_app_infty0073}
 (I) & = & \alpha_n \T{\V{u}} {\nabla} L_n (\V{\beta}_n^0) +
\frac{1}{2} \T{\V{u}} \nabla^2 L_n (\V{\beta}_n^0) \V{u} \alpha_n^2
+ \frac{1}{6} \T{\V{u}} {\nabla} \{ \T{\V{u}} \nabla^2 L_n
(\V{\beta}_n^\ast) \V{u} \} \alpha_n^3 \nonumber \\
 & \triangleq & I_1 + I_2 + I_3,
\end{eqnarray}
where $\V{\beta}_n^\ast$ lies between $\V{\beta}_n^0$ and
$\V{\beta}_n^0 + \alpha_n \V{u}.$ Using the same argument as in the
proof of Theorem 1 of Fan and Peng (2004), we have
\begin{eqnarray}\label{sec_app_infty0073I1}
 |I_1| & = & O_p( \alpha_n^2 n) \| \V{u} \|, \\
 I_2 & = & -\frac{n \alpha_n^2}{2} \T{\V{u}} \M{I}_n (\V{\beta}_n^0) \V{u} +
o_p(1) n \alpha_n^2 \| \V{u} \|^2,
\end{eqnarray}
 and
\begin{eqnarray*}
 | I_3 | & = & \left| \frac{1}{6} \sum_{k_1=1}^{K_n} \sum_{j_1=1}^{p_k}
\sum_{k_2=1}^{K_n} \sum_{j_2=1}^{p_k} \sum_{k_3=1}^{K_n}
\sum_{j_3=1}^{p_k} \frac{\partial^3 L_n(\beta_n^\ast)
}{\partial\beta_{k_{1}j_{1}}\partial\beta_{k_{2}j_{2}}
\partial\beta_{k_{3}j_{3}}} u_{k_{1}j_{1}} u_{k_{2}j_{2}} u_{k_{3}j_{3}} \alpha_n^3 \right|
 \nonumber \\
 & \le & \frac{1}{6} \sum_{i=1}^{n} \left\{ \sum_{k_1=1}^{K_n} \sum_{j_1=1}^{p_k}
\sum_{k_2=1}^{K_n} \sum_{j_2=1}^{p_k} \sum_{k_3=1}^{K_n}
\sum_{j_3=1}^{p_k} M_{{nk_{1}j_{1}}{k_{2}j_{2}}{k_{3}j_{3}}}^{2}
(V_{ni}) \right\}^{1/2}
\| \V{u} \|^3 \alpha_n^3 \nonumber \\
 & = & O_p(P_n^{3/2} \alpha_n ) n \alpha_n^2
\| \V{u} \|^3.
\end{eqnarray*}
 Since
$\frac{ P_n^4 } {n} \rightarrow 0$ and $ P_n^2 \lambda_n \sqrt{a_n }
\rightarrow 0
 $ as $n \rightarrow \infty$, we have
 \begin{equation} \label{sec_app_infty0073I3} | I_3 | = o_p( n \alpha_n^2 )
\| \V{u} \|^3. \end{equation}

From (\ref{sec_app_infty0073I1})-(\ref{sec_app_infty0073I3}), we can
see that, by choosing a sufficiently large $C$, the first term in
$I_2$ dominates $I_1$ uniformly on $\|\V{u}\| = C$; when $n$ is
large enough, $I_2$ also dominates $I_3$ uniformly on $\|\V{u}\| =
C$.

Now we consider $(II)$. Since $\alpha_n = \sqrt{P_n} ( n^{-1/2} +
\lambda_n \sqrt{a_n} /2\sqrt{c_1} ) \rightarrow 0 $, for $\| \V{u}
\| \leq C$ we have
\begin{equation} \label{sec_app_infty:thm1_1}
|\beta_{kj}^0 + \alpha_n u_{kj}
 | \geq |\beta_{kj}^0| - |\alpha_n u_{kj}
 | > 0 \end{equation}
 for $n$ large enough and $\beta_{kj}^0 \neq 0$.
 Hence, we have
\begin{eqnarray*} 
& & p_{\lambda_n,\V{w}_n} ( \V{\beta}_{nk}^0+ \alpha_n \V{u}_k )
 - p_{\lambda_n,\V{w}_n}(\V{\beta}_{nk}^0 )
 \nonumber \\
 & = & \lambda_n ( \sqrt{ w_{n,k1} |\beta_{k1}^0 + \alpha_n u_{k1}
 | + \ldots +w_{n,kp_k} |\beta_{kp_k}^0 + \alpha_n u_{kp_k} | }-
 \sqrt{ w_{n,k1} |\beta_{k1}^0| + \ldots +w_{n,kp_k} |\beta_{kp_k}^0| } ) \nonumber \\
 & \geq & \lambda_n ( \sqrt{ w_{n,k1} |\beta_{k1}^0 + \alpha_n u_{k1}
 | + \ldots + w_{n,ks_k} |\beta_{ks_k}^0 + \alpha_n u_{ks_k} | } -
 \sqrt{ w_{n,k1} |\beta_{k1}^0| + \ldots + w_{n,ks_k} |\beta_{ks_k}^0| } ) \nonumber \\
&\geq & \lambda_n ( \sqrt{ w_{n,k1} |\beta_{k1}^0| +
 \ldots + w_{n,ks_k} |\beta_{ks_k}^0| - \alpha_n( w_{n,k1} | u_{k1}
 | + \ldots + w_{n,ks_k} | u_{ks_k} | )} \nonumber \\
 &&
 ~~~~~~~~~~~~~~ - \sqrt{ w_{n,k1} |\beta_{k1}^0| +
 \ldots + w_{n,ks_k} |\beta_{ks_k}^0| } )
 ~~~~~~~~ (\mathrm{for~ n ~large~ enough, ~ by} ~~ (\ref{sec_app_infty:thm1_1}) )\nonumber \\
 & = & \lambda_n \sqrt{ w_{n,k1} |\beta_{k1}^0| +
 \ldots + w_{n,ks_k} |\beta_{ks_k}^0| } ( \sqrt { 1 - \gamma_{nk} } - 1 ),
 \end{eqnarray*}
where $\gamma_{nk}$ is defined as $\gamma_{nk} =
\frac{\alpha_n(w_{n,k1} | u_{k1} | + \ldots +w_{n,ks_k} | u_{ks_k} |
) }{
 w_{n,k1} |\beta_{k1}^0| + \ldots +w_{n,ks_k} |\beta_{ks_k}^0| } $.
For $n$ large enough, we have $0 \leq \gamma_{nk} < 1$
 and $\gamma_{nk} \leq \frac{\alpha_n \| \V{u}_k \|
 (w_{n,k1} + \ldots +w_{n,ks_k} ) }{
  c_1 ( w_{n,k1} + \ldots +w_{n,ks_k} ) } =
\frac{\alpha_n \| \V{u}_k \| } { c_1 } \leq \frac{\alpha_nC } { c_1
} \rightarrow 0 $ with probability tending to 1 as $n \rightarrow
\infty$.

Therefore,
\begin{eqnarray*} 
& & p_{\lambda_n,\V{w}_n} ( \V{\beta}_{nk}^0+ \alpha_n \V{u}_k )
 - p_{\lambda_n,\V{w}_n}(\V{\beta}_{nk}^0 )
 \nonumber \\
 & \geq & \lambda_n \sqrt{ w_{n,k1} |\beta_{k1}^0| +
 \ldots + w_{n,ks_k} |\beta_{ks_k}^0| } ( \sqrt { 1 - \gamma_{nk} } - 1 ) \nonumber \\
& \geq & \lambda_n \sqrt{ w_{n,k1} |\beta_{k1}^0| +
 \ldots + w_{n,ks_k} |\beta_{ks_k}^0| }
 \left( \frac{1+|o_p(1)|}{2} ( - \gamma_{nk} ) \right) ~~~~~~~~~~~~~~~~~
 \nonumber \\
 &&~~~~~~ ( \mathrm{Using~} \gamma_{nk} = o_p(1) \mathrm{~and ~Taylor ~expansion } )~~~~~~~~~~~~~~~~~
 \nonumber \\
& \geq & - \lambda_n \frac{\alpha_n(w_{n,k1} | u_{k1} | + \ldots
+w_{n,ks_k} | u_{ks_k} | ) }{ \sqrt{
 w_{n,k1} |\beta_{k1}^0| + \ldots +w_{n,ks_k} |\beta_{ks_k}^0| } }
 \left( \frac{1+|o_p(1)|}{2}   \right) \nonumber \\
 & \geq & -\alpha_n \lambda_n
 \frac{ \| \V{u}_k \| \sqrt{a_n s_k}  }{ 2 \sqrt{c_1}  }
 (1+|o_p(1)| ).
\end{eqnarray*}
Therefore, the term $(II)$ in (\ref{sec_app_infty0072a}) is bounded
by
$$ n \alpha_n \lambda_n   \left( \sum_{k=1}^{S_n}
 \frac{ \| \V{u}_k \| \sqrt{a_n s_k}  }{ 2 \sqrt{c_1}  } \right) (1+|o_p(1)|),$$
which is further bounded by
$$ n \alpha_n \lambda_n \sqrt{a_n } (
 \| \V{u} \|  \cdot
\frac{  \sqrt{P_n}  }{ 2 \sqrt{ c_1 } } )  (1+|o_p(1)|).
$$
Note that $\alpha_n = \sqrt{P_n} ( n^{-1/2} + \lambda_n \sqrt{a_n}
/2\sqrt{c_1} ) $, hence the above expression is bounded by
$$ \| \V{u} \| n \alpha_n^2 (1+|o_p(1)|).$$
This term is also dominated by the first term of $I_2$ on $\|\V{u}\|
= C$ uniformly. Therefore, $D_n(\V{u}) < 0$ is satisfied uniformly
on $\|\V{u}\| = C$. This completes the proof of the theorem.

\subsection*{Proof of Theorem 4 }

 We have proved that if $ \lambda_n \sqrt{a_n} = O_p(n^{-1/2} )$,
 there exists a root-$(n/P_n)$ consistent estimate $\hat{\V{\beta}}_n $.
 Now we prove that this root-$(n/P_n)$ consistent estimate has
 the oracle sparsity under the condition
 $ \frac{P_n^{2} }{\lambda_n^2 {b_n}} = o_p( n )$, i.e.,
$\hat{\beta}_{kj} = 0$ with probability tending to 1 if
${\beta}_{kj}^0 = 0$.

 Using Taylor's expansion, we have
\begin{eqnarray} \label{sec_app_infty076}
\frac{\partial Q_n(\V{\beta}_n) }{\partial\beta_{kj}} & = &
\frac{\partial L_n(\V{\beta}_n) }{\partial\beta_{kj}} -
 n \frac{\partial p_{\lambda_n,\V{w}_n}(\V{\beta}_{nk} ) }{\partial\beta_{kj}}
 \nonumber \\
 & = & \frac{\partial L_n(\V{\beta}_n^0) }{\partial\beta_{kj}} +
\sum_{k_{1}=1}^{K_n} \sum_{j_{1}=1}^{p_{k_{1}}}
\frac{\partial^2L_n(\V{\beta}^0)}
{\partial\beta_{kj}\partial\beta_{k_{2}j_{2}} } ( \beta_{k_{1}j_{1}}
- \beta_{k_{1}j_{1}}^0 )
 \nonumber \\
 & & + \frac{1}{2} \sum_{k_{1}=1}^{K_n} \sum_{j_{1}=1}^{p_{k_{1}}}
\sum_{k_{2}=1}^{K_n} \sum_{j_{2}=1}^{p_{k_{2}}}
\frac{\partial^3L_n({\V{\beta}_n^{\ast}})}
{\partial\beta_{kj}\partial\beta_{k_{1}j_{1}}\partial\beta_{k_{2}j_{2}}}
 ( \beta_{k_{1}j_{1}} - \beta_{k_{1}j_{1}}^0 )
 ( \beta_{k_{2}j_{2}} - \beta_{k_{2}j_{2}}^0 )
 \nonumber \\
 & &
- \frac{ n \lambda_n w_{n,kj} }{ 2\sqrt{w_{n,k1} |\beta_{k1}| +
\ldots + w_{n,kp_k} |\beta_{kp_k}| } } \mathrm{sgn} (\beta_{kj})
 \\ \nonumber
 & \triangleq & I_1 + I_2 + I_3 + I_4,
\end{eqnarray}
where ${\V{\beta}_n^{\ast}}$ lies between ${\V{\beta}_n}$ and
$\V{\beta}_n^0$.

Using the argument in the proof of Lemma 5 of Fan and Peng (2004),
for any $\V{\beta}_n$ satisfying $\| {\V{\beta}}_n -\V{\beta}_n^0
 \| = O_p( \sqrt{P_n/n} )$, we have
\begin{eqnarray*}
I_1 & = & O_p(\sqrt{n}) = O_p( \sqrt{nP_n} ), \\
I_2 & = & O_p( \sqrt{nP_n} ), \\
I_3 & = & o_p( \sqrt{nP_n} ).
\end{eqnarray*}

%
%

 Then, since $\hat{\V{\beta}}_n $ is a root-$(n/P_n)$ consistent estimate
 maximizing $Q_n(\V{\beta}_n)$, if $\hat{\beta}_{kj} \neq 0$, we have
\begin{eqnarray} \label{sec_app_infty085}
\left. \frac{\partial Q_n(\V{\beta}_n)
}{\partial\beta_{kj}}\right|_{{\V{\beta}_n} = \hat{\V{\beta}}_n } &
= & O_p( \sqrt{nP_n} ) - \frac{ n \lambda_n w_{n,kj} }{ 2
\sqrt{w_{n,k1} |\hat{\beta}_{k1}| + \ldots + w_{n,kp_k}
|\hat{\beta}_{kp_k}| } } \mathrm{sgn}
(\hat{\beta}_{kj})  \nonumber \\
  & = & 0.
\end{eqnarray}
Therefore, $$ \frac{ n \lambda_n w_{n,kj} }{ \sqrt{ w_{n,k1}
|\hat{\beta}_{k1}| + \ldots + w_{n,kp_k} |\hat{\beta}_{kp_k}| } } =
O_p( \sqrt{nP_n} )~~~~~~\mathrm{for}~\hat{\beta}_{kj} \neq 0. $$
This can be extended to
$$ \frac{ n \lambda_n w_{n,kj}
|\hat{\beta}_{kj}|}{ \sqrt{ w_{n,k1} |\hat{\beta}_{k1}| + \ldots +
w_{n,kp_k} |\hat{\beta}_{kp_k}| } } = |\hat{\beta}_{kj}|
O_p(\sqrt{nP_n} ), $$ for any $\hat{\beta}_{kj}$ with
$\hat{\V{\beta}}_{nk} \neq 0 $. If we sum this over all $j$ in the
$k$th group, we have
\begin{equation}\label{sec_app_infty085a}
n \lambda_n \sqrt{w_{n,k1} |\hat{\beta}_{k1}| + \ldots + w_{n,kp_k}
|\hat{\beta}_{kp_k}| } = \sum_{j=1}^{p_k}|\hat{\beta}_{kj}| O_p(
\sqrt{nP_n} ).
\end{equation}

Since $\hat{\V{\beta}}_n $ is a root-$(n/P_n)$ consistent estimate
of ${\V{\beta}}_n ^0$, we have $|\hat{\beta}_{kj}| = O_p(1)$
 for $(k,j) \in {{\cal{A}}_n } $
and $|\hat{\beta}_{kj}| = O_p( \sqrt{P_n/n} )$
 for $(k,j) \in {\cal{B}}_n \cup {\cal{C}}_n $.


%
%

Now for any $k$ and $j$ satisfying ${\beta}_{kj}^0 = 0$ and
$\hat{\beta}_{kj} \neq 0$, equation (\ref{sec_app_infty085}) can be
written as:
\begin{eqnarray} \label{sec_app_infty086} \left.
\frac{\partial Q_n(\V{\beta}_n)
}{\partial\beta_{kj}}\right|_{{\V{\beta}_n}= \hat{\V{\beta}}_n } & =
&
 \frac{ 1 }{ 2 \lambda_n \sqrt{w_{n,k1} |\hat{\beta}_{k1}|
 + \ldots + w_{n,kp_k} |\hat{\beta}_{kp_k}| } } \\ \nonumber
 & & ( O_p( \sqrt{P_n/n} )
 n \lambda_n \sqrt{w_{n,k1} |\hat{\beta}_{k1}| + \ldots + w_{n,kp_k} |\hat{\beta}_{kp_k}| }
 \\ \nonumber
 & &
 - n \lambda_n^2 w_{n,kj} \mathrm{sgn} (\hat{\beta}_{kj}) ) \\ \nonumber
 & = & 0.
 \end{eqnarray}
Denote $ h_{nk} = O_p( \sqrt{P_n/n} )
 n \lambda_n \sqrt{w_{n,k1} |\hat{\beta}_{k1}| +
 \ldots + w_{n,kp_k} |\hat{\beta}_{kp_k}|  } $.
  Let $ h_n =  \sum_{k=1}^{K_n} h_{nk}   $.
 By equation (\ref{sec_app_infty085a}), we have
 $ h_n =  \sum_{k=1}^{K_n}  O_p( \sqrt{P_n/n} ) \sum_{j=1}^{p_k}|\hat{\beta}_{kj}| O_p(
\sqrt{nP_n} ) = O_p(P_n^2) $. Since $ \frac{ {P_n^{2}} }{\lambda_n^2
{b_n}} = o_p(n)$ guarantees that $n \lambda_n^2 b_n $ dominates
$h_n$ with probability tending to 1 as $n \rightarrow \infty$,
 the first term in (\ref{sec_app_infty086})
 is dominated by the second term as $n \rightarrow \infty$
 uniformly for all $k$ and $j$ satisfying ${\beta}_{kj}^0 = 0$
 since $w_{n,kj} \ge b_n $ and  $h_n > h_{nk}$.
Denote $ g_{nk} =  2\lambda_n \sqrt{w_{n,k1} |\hat{\beta}_{k1}| +
\ldots + w_{n,kp_k} |\hat{\beta}_{kp_k}| } / (n \lambda_n^2 b_n ) $.
Let $ g_n =  \sum_{k=1}^{K_n} g_{nk}   $. By equation
(\ref{sec_app_infty085a}), we have
 $ g_n = 2 \sum_{k=1}^{K_n} (1/n)  \sum_{j=1}^{p_k}|\hat{\beta}_{kj}| O_p(
\sqrt{nP_n} ) /(n \lambda_n^2 b_n ) = o_p(1/\sqrt{nP_n}) $. The
absolute value of the second term in (\ref{sec_app_infty086}) is
bounded below by $1/g_n$. So with probability uniformly converging
to 1 the second term in the derivative $\frac{\partial Q(\V{\beta})
}{\partial \beta_{kj}}|_{\V{\beta} = \hat{\V{\beta}}_n}$ will go to
$\infty$ as $n \rightarrow \infty$, which is a contradiction with
equation (\ref{sec_app_infty086}). Therefore, for any $k$ and $j$
satisfying ${\beta}_{kj}^0 = 0$, we have $\hat{\beta}_{kj} = 0$ with
a probability tending to 1 as $n \rightarrow \infty$. We have $
\hat{\V{\beta}}_{{\cal{D}}_n } = 0 $ with probability tending to 1
as well.

\bigskip

Now we prove the second part of Theorem 4. From the above proof, we
know that there exists $ ( \hat{\V{\beta}}_{n,{{\cal{A}}_n }},
\V{0}) $
 with probability tending to 1, which is a
 root-$(n/P_n)$ consistent local maximizer of
 $Q( {\V{\beta}_n} )$. With a slight abuse of notation, let
$ Q_n(\V{\beta}_{n,{{\cal{A}}_n }} ) = Q_n(\V{\beta}_{n,{\cal{A}}_n
},\V{0} ) $. Using the Taylor expansion on $ \nabla
Q_n(\hat{\V{\beta}}_{n,{{\cal{A}}_n }} )$ at point
$\V{\beta}_{n,{{\cal{A}}_n }}^0$, we have
\begin{eqnarray} \label{sec_app_infty087}
 & &
 \frac{1}{n} ( \nabla^2 L_n( {\V{\beta}}_{n,{{\cal{A}}_n }}^0 )
 ( \hat{\V{\beta}}_{n,{{\cal{A}}_n }} - {\V{\beta}}_{n,{{\cal{A}}_n }}^0 )
 - \nabla J_{n } ( \hat{\V{\beta}}_{n,{{\cal{A}}_n }}) )
 \\ \nonumber
 & = & - \frac{1}{n} \left( \nabla L_n( {\V{\beta}}_{n,{{\cal{A}}_n }}^0 )
 + \frac{1}{2} \T{ ( \hat{\V{\beta}}_{n,{{\cal{A}}_n }} - {\V{\beta}}_{n,{\cal{A}}_n
 }^0)}
 \nabla^2 \{ \nabla L_n( {\V{\beta}}_{n,{{\cal{A}}_n }}^\ast ) \}
 ( \hat{\V{\beta}}_{n,{{\cal{A}}_n }} - {\V{\beta}}_{n,{{\cal{A}}_n }}^0 ) \right)
,
 \end{eqnarray}
where ${\V{\beta}}_{n,{{\cal{A}}_n }}^\ast$ lies between
$\hat{\V{\beta}}_{n,{{\cal{A}}_n }}$ and ${\V{\beta}}_{n,
{\cal{A}}_n }^0$.

Now we define
 $$ {\cal{C}}_n \triangleq \frac{1}{2} \T{ ( \hat{\V{\beta}}_{n,{{\cal{A}}_n }} - {\V{\beta}}_{n,{\cal{A}}_n
 }^0) }
 \nabla^2 \{ \nabla L_n( {\V{\beta}}_{n,{{\cal{A}}_n }}^\ast ) \}
 ( \hat{\V{\beta}}_{n,{{\cal{A}}_n }} - {\V{\beta}}_{n,{{\cal{A}}_n }}^0 ). $$
Using the Cauchy-Schwarz inequality, we have
\begin{eqnarray} \label{sec_app_infty089}
 \left\| \frac{1}{n} {\cal{C}}_n \right\|^2 &
\le &
 \frac{1}{n^2} \sum_{i=1}^n n \| \hat{\V{\beta}}_{n,{{\cal{A}}_n }} - {\V{\beta}}_{n,{{\cal{A}}_n }}^0 \|^4
 \sum_{k_1 =1}^{S_n} \sum_{j_1=1}^{p_k}
\sum_{k_2 =1}^{S_n} \sum_{j_2=1}^{p_k} \sum_{k_3=1}^{S_n}
\sum_{j_3=1}^{p_k} M_{{nk_{1}j_{1}}{k_{2}j_{2}}{k_{3}j_{2}}}^{3}
(\V{V}_{ni})
\nonumber \\
 & = & O_p( {P_n^2}/{n^2} ) O_p(P_n^3) =
 O_p( {P_n^5}/{n^2} ) =  o_p ( {1}/{n} )
. \end{eqnarray}

Since $\frac{ {P_n^5} } {n} \rightarrow 0$ as $n \rightarrow
\infty$, by Lemma 8 of Fan and Peng (2004), we have
$$ \left\| \frac{1}{n} \nabla^2 L_n(
{\V{\beta}}_{n,{{\cal{A}}_n }}^0 ) + \M{I}_n(
{\V{\beta}}_{n,{{\cal{A}}_n }}^0 ) \right\| = o_p ( {1}/{P_n} )
$$
and
\begin{equation}\label{sec_app_infty090} \left\|
\left(\frac{1}{n} \nabla^2 L_n( {\V{\beta}}_{n,{{\cal{A}}_n }}^0 ) +
\M{I}_n( {\V{\beta}}_{n,{{\cal{A}}_n }}^0 ) \right)
 ( \hat{\V{\beta}}_{n,{\cal{A}}_n } - {\V{\beta}}_{n,{\cal{A}}_n }^0 ) \right\| =
 o_p ( {1}/{\sqrt{nP_n}} ) =
 o_p ( {1}/{\sqrt{n }} ).\end{equation}
Since
\begin{eqnarray} \label{appB:crit102}
 & & \sqrt{w_{n,k1} |\hat{\beta}_{k1}| + \ldots + w_{n,ks_k}
|\hat{\beta}_{ks_k}| } \nonumber \\
& = & \sqrt{w_{n,k1} | {\beta}_{k1}^0|( 1+O_p(\sqrt{P_n/n}) ) +
\ldots + w_{n,ks_k} |{\beta}_{ks_k}^0|(
1+O_p(\sqrt{P_n/n}) ) } \nonumber \\
& = &  \sqrt{w_{n,k1} | {\beta}_{k1}^0| + \ldots + w_{n,ks_k}
|{\beta}_{ks_k}^0| } ( 1+O_p(\sqrt{P_n/n}) ) \nonumber,
\end{eqnarray}
 we have
 $$ \frac{
\lambda_n w_{n,kj} }{ \sqrt{w_{n,k1} |\hat{\beta}_{k1}| + \ldots +
w_{n,ks_k} |\hat{\beta}_{ks_k}| } } = \frac{ \lambda_n w_{n,kj} }{
\sqrt{w_{n,k1} | {\beta}_{k1}^0| + \ldots + w_{n,ks_k}
|{\beta}_{ks_k}^0| } } ( 1+O_p(\sqrt{P_n/n}) ) .$$
 Furthermore, since $$ \frac{
\lambda_n w_{n,kj} }{ \sqrt{w_{n,k1} | {\beta}_{k1}^0| + \ldots +
w_{n,ks_k} |{\beta}_{ks_k}^0| } } \le \frac{ \lambda_n w_{n,kj} }{
\sqrt{ w_{n,kj} c_1 } }
 \leq \frac{ \lambda_n \sqrt{ a_n }
 }{ \sqrt{ c_1 } }
= o_p( {(nP_n)}^{-1/2} ) $$ for $ (k,j) \in {{\cal{A}}_n } $, we
have
$$ \left( \frac{1}{n} \nabla J_{n } (
\hat{\V{\beta}}_{n,{\cal{A}}_n }) \right)_{kj} = \frac{ \lambda_n
w_{n,kj} }{ 2 \sqrt{w_{n,k1} |\hat{\beta}_{k1}| + \ldots +
w_{n,ks_k} |\hat{\beta}_{ks_k}| } } = o_p( {(nP_n)}^{-1/2} ) $$ and
\begin{equation}\label{sec_app_infty091}
 \left\| \frac{1}{n} \nabla J_{n } (
\hat{\V{\beta}}_{n,{\cal{A}}_n }) \right\| \le \sqrt{P_n} o_p(
{(nP_n)}^{-1/2} ) = o_p ( {1}/{\sqrt{n }} ).
\end{equation}

Together with (\ref{sec_app_infty089}), (\ref{sec_app_infty090}) and
(\ref{sec_app_infty091}), from (\ref{sec_app_infty087}) we have
$$ \M{I}_n( {\V{\beta}}_{n,{{\cal{A}}_n }}^0 )
 ( \hat{\V{\beta}}_{n,{{\cal{A}}_n }} - {\V{\beta}}_{n,{{\cal{A}}_n }}^0 )
 = \frac{1}{n} \nabla L_n( {\V{\beta}}_{n,{{\cal{A}}_n }}^0 )
 + o_p( {1}/{\sqrt{n }} ).$$

Now using the same argument as in the proof of Theorem 2 of Fan and
Peng (2004), we have
\begin{equation*}
 \sqrt{n} \M{A}_n \M{I}_n^{1/2} ( {\V{\beta}}_{n,{{\cal{A}}_n }}^0 )
 ( \hat{\V{\beta}}_{n,{{\cal{A}}_n }} - {\V{\beta}}_{n,{{\cal{A}}_n }}^0
 ) \rightarrow \sqrt{n} \M{A}_n \M{I}_n^{-1/2} ( {\V{\beta}}_{n,{{\cal{A}}_n }}^0 )
 \left(\frac{1}{n} \nabla L_n( {\V{\beta}}_{n,{{\cal{A}}_n }}^0 )\right)
 \rightarrow {\cal{N}} (\V{0}, \M{G}),
 \end{equation*}
where $ \M{A}_n $ is a $q \times |{{\cal{A}}_n }| $ matrix such that
$ \M{A}_n \T{\M{A}_n} \rightarrow \M{G}$ and $\M{G}$ is a $q\times
q$ nonnegative symmetric matrix.

\subsection*{Proof of Theorem 2}

Note that when $w_{n,kj} = 1 $, we have $a_n = 1$ and $b_n = 1$. The
conditions $ \lambda_n \sqrt{a_n } = O_p(n^{-1/2} )$ and
 $ \frac{ {P_n^{2}} }{\lambda_n^2 {b_n}} = o_p(n )$ in Theorem
 4 become $\lambda_n \sqrt{n} = O_p(1) $ and $\frac{P_n}{\lambda_n
\sqrt{n} } \rightarrow 0$. These two conditions cannot be satisfied
simultaneously by adjusting $\lambda_n$, which implies that
$\F{Pr}(\hat{\V{\beta}}_{\cal{D}} = 0) \rightarrow 1 $ cannot be
guaranteed.

We will prove that by choosing $\lambda_n$ satisfying
$\sqrt{n}{\lambda_n} = O_p(1) $ and $P_n n^{-3/4}/{\lambda_n}
\rightarrow 0$ as $n \rightarrow \infty$, we can have a root-$n$
consistent local maximizer $\hat{\V{\beta}}_n =
 \T{(\hat{\V{\beta}}_{{\cal A}_n}, \hat{\V{\beta}}_{{\cal B}_n},
 \hat{\V{\beta}}_{{\cal C}_n})}$ such that
$\F{Pr}(\hat{\V{\beta}}_{{\cal C}_n} = 0) \rightarrow 1$.

 Similar as
in the proof of Theorem 4, we let
  $ h_n^{\prime} =  \sum_{k=S_n+1}^{K_n} h_{nk}   $.
 By equation (\ref{sec_app_infty085a}), we have
 $ h_n^{\prime} =  \sum_{k=S_n+1}^{K_n}  O_p( \sqrt{P_n/n} )
 \sum_{j=1}^{p_k}|\hat{\beta}_{kj}| O_p(
\sqrt{nP_n} ) = O_p(P_n^2/\sqrt{n}) $. Since $P_n
n^{-3/4}/{\lambda_n} \rightarrow 0$ guarantees that $n \lambda_n^2 $
dominates $h_n^{\prime}$ with probability tending to 1 as $n
\rightarrow \infty$,
 the first term in (\ref{sec_app_infty086})
 is dominated by the second term as $n \rightarrow \infty$
 uniformly
 for any $k$ satisfying ${\V{\beta}}_{nk}^0 = 0$
 since $w_{n,kj} = 1 $ and  $h_n^{\prime} > h_{nk}$.
Similar as in the proof of Theorem 4, we have $
\hat{\V{\beta}}_{{\cal{C}}_n } = 0 $ with probability tending to 1.

\subsection*{Proof of Theorem 5}

Let $ N_n = |{{\cal{A}}_n }| $ be the number of nonzero parameters.
  Let $\M{B}_n$ be an $( N_n - q ) \times N_n $ matrix which satisfies
$\M{B}_n \T{\M{B}}_n = \M{I}_{N_n - q } $ and $\M{A}_n \T{\M{B}}_n =
0 $. As ${\V{\beta}}_{n,{{\cal{A}}_n }}$ is in the orthogonal
complement to the linear space that is spanned by the rows of
$\M{A}_n$ under the null hypothesis $H_0$, it follows that
 $$ {\V{\beta}}_{n,{{\cal{A}}_n }} =  \T{\M{B}}_n {\V{\gamma}}_{n}  , $$
where ${\V{\gamma}}_{n}$ is an $( N_n - q ) \times 1 $ vector. Then,
under $H_0$ the penalized likelihood estimator is also the local
maximizer $\hat{\V{\gamma}}_{n}$ of the problem

$$ Q_n({\V{\beta}}_{n,{{\cal{A}}_n }}) = \max_{ {\V{\gamma}}_{n} } Q_n( \T{\M{B}}_n {\V{\gamma}}_{n} ). $$

To prove Theorem 5 we need the following two lemmas.

\begin{lemma}
Under condition $(b)$ of Theorem 4 and the null hypothesis $H_0$, we
have
$$     \hat{\V{\beta}}_{n,{{\cal{A}}_n }} - {\V{\beta}}_{n,{{\cal{A}}_n }}^0
   =  \frac{1}{n} \M{I}_n^{-1} ( {\V{\beta}}_{n,{{\cal{A}}_n }}^0 )
  \nabla L_n( {\V{\beta}}_{n,{{\cal{A}}_n }}^0 )  + o_p(n^{-1/2}), $$
$$  \T{\M{B}}_n ( \hat{\V{\gamma}}_{n} - {\V{\gamma}}_{n}^0
 ) =  \frac{1}{n} \T{\M{B}}_n \{ \M{B}_n \M{I}_n( {\V{\beta}}_{n,{{\cal{A}}_n }}^0 ) \T{\M{B}}_n \}^{-1}
 \M{B}_n \nabla L_n( {\V{\beta}}_{n,{{\cal{A}}_n }}^0 )  + o_p(n^{-1/2}). $$
\end{lemma}

\subsection*{Proof of of Lemma 3}

We need only prove the second equation. The first equation can be
shown in the same manner. Following the proof of Theorem 4, it
follows that under $H_0$,

$$ \M{B}_n \M{I}_n( {\V{\beta}}_{n,{{\cal{A}}_n }}^0 ) \T{\M{B}}_n  ( \hat{\V{\gamma}}_{n} -
{\V{\gamma}}_{n}^0  ) = \frac{1}{n} \M{B}_n \nabla L_n(
{\V{\beta}}_{n,{{\cal{A}}_n }}^0 ) + o_p(n^{-1/2}).
  $$
As the eigenvalue $\lambda_i ( \M{B}_n \M{I}_n(
{\V{\beta}}_{n,{{\cal{A}}_n }}^0 ) \T{\M{B}}_n ) $ is uniformly
bounded away from 0 and infinity, we have
$$  \T{\M{B}}_n ( \hat{\V{\gamma}}_{n} - {\V{\gamma}}_{n}^0
 ) =  \frac{1}{n} \T{\M{B}}_n \{ \M{B}_n \M{I}_n( {\V{\beta}}_{n,{{\cal{A}}_n }}^0 ) \T{\M{B}}_n \}^{-1}
 \M{B}_n \nabla L_n( {\V{\beta}}_{n,{{\cal{A}}_n }}^0 )  + o_p(n^{-1/2}). $$

\begin{lemma}
Under condition $(b)$ of Theorem 4 and the null hypothesis $H_0$, we
have
\begin{eqnarray} \label{sec_app_infty095}
 & &
 Q_n(\hat{\V{\beta}}_{n,{{\cal{A}}_n }}) - Q_n( \T{\M{B}}_n \hat{\V{\gamma}}_{n} )
 \\   \nonumber
 & = & \frac{n}{2} \T{ ( \hat{\V{\beta}}_{n,{{\cal{A}}_n }}  -  \T{\M{B}}_n \hat{\V{\gamma}}_{n} ) }
  \M{I}_n ( {\V{\beta}}_{n,{{\cal{A}}_n }}^0 )
  { ( \hat{\V{\beta}}_{n,{{\cal{A}}_n }}  -  \T{\M{B}}_n \hat{\V{\gamma}}_{n} ) } + o_p(1).
 \end{eqnarray}
\end{lemma}

\subsection*{Proof of Lemma 4}

 A Taylor's expansion of $ Q_n(\hat{\V{\beta}}_{n,{{\cal{A}}_n
}}) - Q_n( \T{\M{B}}_n \hat{\V{\gamma}}_{n} )$ at the point $
\hat{\V{\beta}}_{n,{{\cal{A}}_n }} $ yields
$$  Q_n(\hat{\V{\beta}}_{n,{{\cal{A}}_n }}) - Q_n( \T{\M{B}}_n \hat{\V{\gamma}}_{n} ) = T_1 + T_2 + T_3 + T_4,$$
where
\begin{eqnarray*}
 T_1 & =  & \T{\nabla} Q_n(\hat{\V{\beta}}_{n,{{\cal{A}}_n }})
     { ( \hat{\V{\beta}}_{n,{{\cal{A}}_n }}  -  \T{\M{B}}_n \hat{\V{\gamma}}_{n} ) } , \\
 T_2 & =  &  - \frac{1}{2} \T{ ( \hat{\V{\beta}}_{n,{{\cal{A}}_n }}  -  \T{\M{B}}_n \hat{\V{\gamma}}_{n} ) }
  \nabla^2 L_n( \hat{\V{\beta}}_{n,{{\cal{A}}_n }} )
  { ( \hat{\V{\beta}}_{n,{{\cal{A}}_n }}  -  \T{\M{B}}_n \hat{\V{\gamma}}_{n} ) } , \\
 T_3 & =  &    \frac{1}{6} \T{\nabla}
   \{
        \T{ ( \hat{\V{\beta}}_{n,{{\cal{A}}_n }}  -  \T{\M{B}}_n \hat{\V{\gamma}}_{n} ) }
      \nabla^2 L_n(  {\V{\beta}}_{n,{{\cal{A}}_n }}^{\star} )
      { ( \hat{\V{\beta}}_{n,{{\cal{A}}_n }}  -  \T{\M{B}}_n \hat{\V{\gamma}}_{n} ) }
    \}
    { ( \hat{\V{\beta}}_{n,{{\cal{A}}_n }}  -  \T{\M{B}}_n \hat{\V{\gamma}}_{n} ) } , \\
 T_4 & =  &  \frac{1}{2} \T{ ( \hat{\V{\beta}}_{n,{{\cal{A}}_n }}  -  \T{\M{B}}_n \hat{\V{\gamma}}_{n} ) }
  \nabla^2 J_{n} (  {\V{\beta}}_{n,{{\cal{A}}_n }}^{\ast} )
  { ( \hat{\V{\beta}}_{n,{{\cal{A}}_n }}  -  \T{\M{B}}_n \hat{\V{\gamma}}_{n} )
  } .
 \end{eqnarray*}

We have $ T_1 = 0 $ as $ \T{\nabla}
Q_n(\hat{\V{\beta}}_{n,{{\cal{A}}_n }}) = 0$.

Let $ \M{\Theta}_n = \M{I}_n  ( {\V{\beta}}_{n,{{\cal{A}}_n }}^0 )$
and $ \M{\Phi}_n = \frac{1}{n}   \nabla L_n(
{\V{\beta}}_{n,{{\cal{A}}_n }}^0 ) $. By Lemma 2 we have
\begin{eqnarray*}
  &   &   { ( \hat{\V{\beta}}_{n,{{\cal{A}}_n }}  -  \T{\M{B}}_n \hat{\V{\gamma}}_{n} ) }
    \nonumber \\
 & = & \M{\Theta}_n^{-1/2} \{
    \M{I}_n -  \M{\Theta}_n^{1/2} \T{\M{B}}_n ( \M{B}_n \M{\Theta}_n \T{\M{B}}_n )^{-1} \M{B}_n\M{\Theta}_n^{1/2}
 \}
 \M{\Theta}_n^{-1/2} \M{\Phi}_n  \nonumber \\
 &   &   +    o_p(n^{-1/2}).
 \end{eqnarray*}
  $\M{I}_n -  \M{\Theta}_n^{1/2} \T{\M{B}}_n ( \M{B}_n \M{\Theta}_n \T{\M{B}}_n )^{-1} \M{B}_n\M{\Theta}_n^{1/2}$ is an
  idempotent matrix with rank $q$. Hence, by a standard argument and condition
 (A2),
 $$  ( \hat{\V{\beta}}_{n,{{\cal{A}}_n }}  -  \T{\M{B}}_n \hat{\V{\gamma}}_{n} )  = O_p( \sqrt{\frac{q}{n} } ) .$$

We have \begin{equation} \label{sec_app_infty096} \left( \frac{1}{n}
\nabla^2 J_{n } ( {\V{\beta}}_{n,{\cal{A}}_n }) \right)_{kjk_1j_1} =
0, ~~~~~~\mathrm{for}~k \ne k_1 ~~~~~~
\end{equation} and
\begin{eqnarray} \label{sec_app_infty097}
  &   &   \left( \frac{1}{n} \nabla^2 J_{n } (
 {\V{\beta}}_{n,{\cal{A}}_n }^{\ast} ) \right)_{kjkj_1}   \nonumber \\
 & =  & \frac{ \lambda_n
w_{n,kj} w_{n,kj_1}  }{ 4 ( {w_{n,k1} | {\beta}_{k1}^{\ast} | +
\ldots + w_{n,ks_k} | {\beta}_{ks_k}^{\ast} | } )^{3/2} } \nonumber \\
 & =  & \frac{ \lambda_n
w_{n,kj} w_{n,kj_1}  }{ 4 ( {w_{n,k1} | {\beta}_{k1}^{0} | + \ldots
+ w_{n,ks_k} | {\beta}_{ks_k}^{0} | } )^{3/2} } (1+o_p(1) ) \nonumber \\
 & \le  & \frac{ \lambda_n \sqrt{a_n}  }{ 4 ( c_1 )^{3/2} } (1+o_p(1) ) \nonumber \\
 & =  &  o_p( (nP_n)^{-1/2} ).
 \end{eqnarray}
Combining (\ref{sec_app_infty096}), (\ref{sec_app_infty097}) and
condition $ q < P_n $, following the proof of $I_3$ in Theorem 3, we
have
   $$ T_3 = O_p( n P_n^{3/2} n^{-3/2} q ^{3/2}  )  = o_p(1) $$
   and
 \begin{eqnarray*}
 T_4  &  \le &   n  \left\| \frac{1}{n} \nabla^2 J_{n } (
 {\V{\beta}}_{n,{\cal{A}}_n }^{\ast} ) \right\|
 \| \hat{\V{\beta}}_{n,{{\cal{A}}_n }}  -  \T{\M{B}}_n \hat{\V{\gamma}}_{n} \|^2
   \nonumber \\
 & =  & n {P_n} o_p( (nP_n)^{-1/2} ) O_p(\frac{q}{n})
 \nonumber \\
 & =  &  o_p(1).
 \end{eqnarray*}
    Thus,
\begin{equation} \label{sec_app_infty098}
   Q_n(\hat{\V{\beta}}_{n,{{\cal{A}}_n }}) - Q_n( \T{\M{B}}_n \hat{\V{\gamma}}_{n} ) =
       T_2 + o_p(1).
\end{equation}
It follows from Lemmas 8 and 9 of Fan and Peng (2004) that
  $$  \left\| \frac{1}{n} \nabla^2 L_n( \hat{\V{\beta}}_{n,{{\cal{A}}_n }} )  +
     \M{I}_n ( {\V{\beta}}_{n,{{\cal{A}}_n }}^0 )  \right\| =
     o_p\left( \frac{1}{\sqrt{P_n}} \right).$$
Hence, we have
\begin{eqnarray} \label{sec_app_infty099}
  &   &    \frac{1}{2} \T{ ( \hat{\V{\beta}}_{n,{{\cal{A}}_n }}  -  \T{\M{B}}_n \hat{\V{\gamma}}_{n} ) }
  \{
  \nabla^2 L_n( \hat{\V{\beta}}_{n,{{\cal{A}}_n }} )
 + n \M{I}_n ( {\V{\beta}}_{n,{{\cal{A}}_n }}^0 )
 \}
  { ( \hat{\V{\beta}}_{n,{{\cal{A}}_n }}  -  \T{\M{B}}_n \hat{\V{\gamma}}_{n} ) }  \nonumber \\
 & \le  & o_p\left( n\frac{1}{\sqrt{P_n}} \right) O_p( {\frac{q}{n} } ) =
    o_p(1).
 \end{eqnarray}
 The combination of (\ref{sec_app_infty098}) and (\ref{sec_app_infty099}) yields (\ref{sec_app_infty095}).

\subsection*{Proof of Theorem 5 }

Given Lemmas 3 and 4, the proof of the Theorem is similar to the
proof of Theorem 4 in Fan and Peng (2004).

\end{document}